\documentstyle[12pt]{article}

\renewcommand{\theequation}{15.\arabic{equation}}
\newcommand{\SV}{S_{\mbox{\scriptsize vac}}}
\newcommand{\SC}{S_{\mbox{\scriptsize cl}}}
\newcommand{\SS}{S_{\mbox{\scriptsize source}}}
\newcommand{\ja}{j^{\alpha}_{\,\mbox{\scriptsize bare}}}
\newcommand{\jm}{j^{\mu}_{\,\mbox{\scriptsize bare}}}
\newcommand{\E}{{\cal E}}
\newcommand{\e}{{\rm\bf e}}
\newcommand{\be}{{\rm\bf e}_{\;\mbox{\scriptsize bare}}}
\newcommand{\r}{r_{\mbox{\scriptsize min}}}
\renewcommand{\t}{t_{\mbox{\scriptsize start}}}
\newcommand{\rol}{\rho_{\,\mbox{\scriptsize lim}}}
\newcommand{\F}{F_{\mbox{\scriptsize stat}}}
\newcommand{\R}{\hat{\cal R}}
\renewcommand{\H}{{\cal H}_q}
\newcommand{\tr}{{\mbox{tr}}}
\newcommand{\integral}{\int dx\, g^{1/2}}
\newcommand{\bintegral}{\int d{\bar x}\, {\bar g}^{1/2}}
\newcommand{\sint}{\int\limits^{\infty}_{4m^2} d\mu^2\,\Gamma (\mu^2)\,}
\newcommand{\bes}{\int\limits^0_{-\infty} dq\, J_0 (\mu\sqrt{-2q})}
\newcommand{\cint}{
\int\limits_{\mbox{\scriptsize past of}\; x} d{\bar x}\,{\bar g}^{1/2}
}
\newcommand{\phint}{
\int\limits_{\mbox{\scriptsize past of}\; x} d{\bar t}\,d{\bar r}\,
}
\makeatletter
\@addtoreset{equation}{section}
\makeatother
\begin{document}
{\renewcommand{\theequation}{1.\arabic{equation}}

\begin{center}
{\LARGE\bf 
On electrodynamics of rapidly\\ moving sources}
\end{center}
\begin{center}
{\bf R. Pettorino}
\end{center}
\begin{center}
Dipartimento di Scienze Fisiche, Universit\`a di Napoli,
and INFN - Sezione di Napoli,\\ Monte S. Angelo, via Cinthia,
I-80126 Naples, Italy
\end{center}
\begin{center}
and
\end{center}
\begin{center}
{\bf G.A. Vilkovisky}
\end{center}
\begin{center}
Lebedev Physics Institute and
Research Center in Physics,\\ Leninsky Prospect 53, Moscow 117924,
Russia 
\end{center}
\vspace{2cm}
\begin{abstract}

Rapidly moving sources create pairs in the vacuum
and lose energy. In consequence of this, the velocity of a charged
body cannot approach the speed of light closer than a certain
limit which depends only on the coupling constant. The vacuum 
back-reaction secures the observance of the conservation laws.
A source can lose up to 50\% of energy and charge because of 
the vacuum instability.
\end{abstract}
\newpage

\begin{center}
\section{\bf    Introduction} 
\end{center}

The present article is an extended account of the work reported
in Ref. [1]. We consider a classical system of electric charges
which make a source of the electromagnetic field and move in
the self field. However, we take into account that this source is
immersed in the real vacuum, and the field that it generates excites 
the vacuum charges. The problem is figuring out the vacuum
back-reaction on the motion of the source.

The electromagnetic field generated by a source is a solution of
the expectation-value equations in the in-vacuum state. For this
state to exist [2], the source is assumed asymptotically static
in the past. In consequence of this assumption, the solution always
contains a contribution of the static vacuum polarization whose
principal effect is screening of the monopole moment of the
source. The polarization occurs in the whole of space and increases
infinitely in the vicinity of the source thereby causing the
ultraviolet disaster. However, we show that this infinity does not
affect the motion of the source. After the self-action of a pointlike
charge is properly eliminated, the force exerted by the source on itself
is finite. By choosing the spatial scale of the system exceeding the
Compton size of the vacuum particle, we abstract ourselves from
the static polarization altogether.

The effect that we are concerned about is the vacuum instability
caused by a nonstationarity of the source [2,3]. A nonstationary
electromagnetic field is capable of creating in the vacuum real
particles having the electric charge. When the frequency of the
source exceeds the threshold of pair creation, it emits a flux
of energy and charge carried by the created particles. The main
question is how much energy can be extracted from a source by
means of this mechanism? An attempt to answer this question
without taking into account the back-reaction of the vacuum on
the motion of the source leads to a contradiction with the
energy conservation law [3]. The radiation rate grows unboundedly
with the energy of the source, and, at a sufficiently high energy,
the source appears to give more than it has.

The problem of self-consistent motion is solved below for the
simplest model of a pair-creating source. The model is a charged
spherical shell expanding in the self field. This choice is made
to avoid the complications connected with an emission of the
electromagnetic waves. A high-frequency source will generally emit
both the electromagnetic waves by the classical mechanism and
charged particles by the quantum mechanism. The two radiations
overlap in a nontrivial way: the energy of the vacuum of charged
particles goes partially into the electromagnetic radiation and
amplifies it [4]. Accounting for the vacuum back-reaction is then
necessary already for a removal of the infrared disaster, and the
problem of restoration of the energy conservation law concerns both
components of radiation [4]. By choosing the source spherically
symmetric, we exclude the emission of waves, and, thereby, put off
the solution of this more complicated problem.

The solution in the case of the spherical shell is in the fact that
there emerges a new kinematic bound on the velocity of the source.
Raising the energy of the source results in an increase of its
acceleration, which causes an intensification of the vacuum particle
production, which entails a reinforcement of its back-reaction, which 
results in a deceleration of the source. As a result, however high
the energy may be, the velocity of a charged body cannot approach
the speed of light closer than a certain limit. Within a given type
of coupling, this limit is universal. It does not depend on the
parameters of the source, only on the coupling constant. The 
back-reaction effect is nonanalytic in the coupling constant and restores
completely the conservation laws. Up to 50\% of energy and charge can
be extracted from the source by raising its initial energy.

The effect of vacuum instability is of significance for rapidly
moving, or high-frequency sources. The high-frequency approximation
[3,4] is the only approximation made in the solution. The 
phenomenological, or axiomatic theory of the vacuum [5] is used in which 
the expectation-value equations are specified by a set of operator
form factors. In the high-frequency approximation, only the polarization
operator is involved in the calculation of the induced 
charge\footnote{For the calculation of the induced energy, one needs
also the gravitational form factors [3,4].}. Its relevant properties
are postulated in Section 2.

In Section 3, a closed set of equations is obtained for the motion
of the source in the self electromagnetic field. The technique needed
for dealing with the nonlocal expectation-value equations is presented
in Section 4. In Section 5, the static solution is considered, valid
outside some future light cone. The solution for a moving source is
obtained in Section 6, and its ultraviolet behaviour is studied in
Sections 7 and 8. In Sections 9-11, the force of the vacuum 
back-reaction is calculated, and the equation of motion of the source 
is solved in the high-frequency approximation. The rate of emission 
of charge is calculated in Section 12.
}

\newpage
{\renewcommand{\theequation}{2.\arabic{equation}}

\begin{center}
\section{\bf    Electrically charged source coupled to the
vacuum charges}
\end{center}

Our starting point is the action for the electromagnetic field
generated by a source
\begin{equation}
S=\SC + \SV + \SS
\end{equation}
in which the source is a set of particles with masses $M_i$ and
charges $e_i$ :
\begin{equation}
\SS =\sum_i \int ds\,\left(\frac{M_i}{2} g_{\mu\nu}\left(x_i(s)\right)
\frac{dx^\mu_i}{ds}\frac{dx^\nu_i}{ds}+e_i A_\mu\left(x_i(s)\right)
\frac{dx^\mu_i}{ds}\right)\; .
\end{equation}
Here $g_{\mu\nu}$ is the flat metric
\begin{equation}
g_{\mu\nu}dx^\mu dx^\nu =-dt^2+dr^2+r^2(d\theta^2 +\sin^2\theta\,
d\varphi^2)\; ,
\end{equation}
$A_\mu (x)$ is the electromagnetic potential, and $s$ is the proper
time of the $i$-th particle. The quantities
\begin{equation}
M=\sum_i M_i\qquad ,\qquad e=\sum_i e_i
\end{equation}
are the full mass and charge of the source ($c=1$). For definiteness,
the charge $e$ will be considered positive.

In Eq. (2.1), $\SC$ is the classical action of the electromagnetic field
\begin{equation}
\SC =-\frac{1}{16\pi}\integral F_{\mu\nu}F^{\mu\nu}\qquad ,\qquad
F_{\mu\nu}=\partial_\mu A_\nu -\partial_\nu A_\mu\; ,
\end{equation}
and $\SV$ is the effective action of the vacuum charges. The 
phenomenological, or axiomatic theory of the vacuum [5] will be used
in which $\SV$ is taken as the general expansion over the basis of
nonlocal invariants [6]:
\begin{equation}
\SV =-\frac{1}{16\pi}\integral F_{\mu\nu}f(-\Box)F^{\mu\nu}+
O(F\times F\times F\ldots)\; .
\end{equation}
The higher-order terms of this expansion are of the form
\begin{equation}
\integral f(\Box_1,\Box_2,\Box_3,\ldots\Box_{1+2},\Box_{1+3},\ldots)
F_1\times F_2\times F_3\ldots
\end{equation}
where $f$'s are functions of the D'Alembert operators (the form factors).
In (2.7), the operator $\Box_n$ acts on $F_n$ , and the operator
$\Box_{m+n}$ acts on the product $F_m F_n$ . All form factors are
assumed to admit spectral representations through the resolvents
$1/(\mu^2 -\Box )$ .

The expectation-value equations of the electromagnetic field, 
associated\footnote{For the expectation-value equations there is
no direct least-action principle but they differ from the
variational equations of the Feynman effective action only by the
boundary conditions for the resolvents [7].} with the action (2.1)
are the following equations for the current $j^\alpha$ :
\begin{equation}
\nabla_\beta F^{\alpha\beta}=4\pi j^\alpha\; ,
\end{equation}
\begin{equation}
j^\alpha + f(-\Box)j^\alpha + O(F\times F\ldots)=\ja
\end{equation}
with the retarded resolvents [5] for $f(-\Box)$ and the higher-order
form factors. The current $\ja$ as given by the action (2.2) is of
the form
\begin{equation}
\ja (x)=\sum_i \int ds\, e_i\,\frac{1}{g^{1/2}(x)}\delta^{(4)}
\left( x-x_i (s)\right)\frac{dx^\alpha_i (s)}{ds}\; .
\end{equation}

The full set of exact form factors provides a complete phenomenological 
description of the vacuum of particles having a given type of charge
(here the electric charge). On the other hand, the form factors are
to be calculated from some quantum-field model, and, even within a given
model, this calculation can never be complete let alone the fact
that one never knows the ultimate model of the vacuum. The virtue
of the axiomatic approach is in the fact that, for the physical
questions of interest, the detailed form of the form factors is
unimportant. Only some of their properties are important. These
properties should be postulated, and the expectation-value problem
solved with the form factors which are otherwise arbitrary. The
axiomatic approach is model-independent, and at the same time it is
a tool for testing various models and approximations therein. [5]
The present paper gives an example of this approach.

A possibility of truncating the series (2.6) depends on the
expectation-value problem in question which, in its turn, is specified
by the properties of the source\footnote{We assume that there are no
incoming electromagnetic waves.}. Our concern in the present paper is
a high-frequency source creating pairs. Let $\nu$ be its typical
frequency. In the problem of particle creation by a nonstationary
external or mean field, this field is considered high-frequency if
the energy $\hbar\nu$ dominates both the rest energy of the vacuum 
particle and its static (Coulomb) energy in this field [3,4].
The high-frequency approximation is the condition of validity of
the expansion (2.6) [3].

The linear expectation-value equations obtained by truncating
Eq. (2.9) are solved by the ansatz
\begin{equation}
j^\alpha = \ja -\gamma (-\Box)\ja
\end{equation}
in which $\gamma (-\Box)$ is some retarded form factor. Its relevant
properties are to be postulated. We assume that the function
$\gamma (-\Box)$ is analytic in the complex plane of $-\Box$ except
at the real negative half-axis where it has a cut:
\begin{equation}
\frac{1}{2\pi{\rm i}}[\gamma(-\mu^2 -{\rm i}0) - 
\gamma(-\mu^2 +{\rm i}0)]=\Delta(\mu^2)\; .
\end{equation}
The properties of the spectral-mass function $\Delta(\mu^2)$ that 
need to be specified are (i) positivity
\begin{equation}
\Delta(\mu^2)\ge 0\; ,
\end{equation}
(ii) the presence of a lower bound in the spectrum
\begin{equation}
\Delta(\mu^2)\propto\theta (\mu^2 -4m^2)\; ,\qquad m\ne 0\; ,
\end{equation}
and (iii) finiteness at large spectral mass
\begin{equation}
\Delta(\mu^2)\Bigl|_{\mu^2\to\infty}=\frac{\kappa^2}{24\pi}\ne 0\; .
\end{equation}
Eq. (2.14) introduces the mass of the vacuum particles $m$ , and
Eq. (2.15) introduces the coupling constant $\kappa^2$ . Finally,
the function $\gamma (-\Box)$ must satisfy the normalization condition
\begin{equation}
\gamma (0)=0\; .
\end{equation}

Redefining the spectral-mass function as
\begin{equation}
\Delta (\mu^2)=\frac{\kappa^2}{24\pi}\Gamma (\mu^2)\; ,\qquad 
\mu^2\ge 4m^2\; ,
\end{equation}
we obtain from Eqs. (2.12)-(2.16)
\begin{equation}
\gamma (-\Box)=\frac{\kappa^2}{24\pi}\sint\left(\frac{1}{\mu^2 -\Box}
-\frac{1}{\mu^2}\right)\; .
\end{equation}
To summarize, we assume that the expectation-value equations (2.11)
hold with the form factor (2.18) in which the resolvent
$1/(\mu^2 -\Box )$ is retarded [5,7], and $\Gamma (\mu^2)$ satisfies
the conditions
\begin{equation}
\Gamma (\mu^2)\ge 0\qquad ,\qquad \Gamma (\infty)=1\; .
\end{equation}

In the underlying quantum field theory, Eqs. (2.13)-(2.15) assume
(i) positivity of the metric of the physical Hilbert space,
(ii) the presence of an energy threshold for pair creation, and
(iii) a logarithmic divergence of the charge renormalization.
Eq. (2.16) is a condition that $\kappa^2$ is the renormalized
coupling constant. For example, in the case of the electron-positron 
vacuum, the equations above hold with
\begin{equation}
\Gamma (\mu^2)=\Bigl(1-\frac{4m^2}{\mu^2}\Bigr)^{1/2}
\Bigl(1+\frac{2m^2}{\mu^2}\Bigr)+\mbox{ multi-loop contributions}
\end{equation}
and
\begin{equation}
\kappa^2=8\alpha + O(\alpha^2)
\end{equation}
where $\alpha$ is the fine-structure constant. In the case where
$\SV$ is the standard loop [3] with the abelian commutator 
curvature\footnote{For the standard loop with arbitrary metric, 
connection, and potential the calculations can be carried out
in the general form, and the results tabulated. The one-loop
action for any model is then obtained by combining the standard
loops. (See Refs. [3,4] and references therein.)}
\begin{equation}
\R_{\mu\nu}={\hat\Omega}F_{\mu\nu}\; ,
\end{equation}
the coupling constant is defined by the matrix ${\hat\Omega}$ :
\begin{equation}
\kappa^2 =-\tr\,{\hat\Omega}^2\; ,
\end{equation}
and the spectral-mass function is
\begin{equation}
\Gamma (\mu^2)=\Bigl(1-\frac{4m^2}{\mu^2}\Bigr)^{3/2}\; .
\end{equation}
}

\newpage
{\renewcommand{\theequation}{3.\arabic{equation}}

\begin{center}
\section{\bf    The charged shell expanding in the self field}
\end{center}

To avoid complications connected with an emission of the electromagnetic
waves [4], the source will be chosen spherically symmetric, and,
moreover, the particles in the action (2.2) will be assumed to pack
a thin spherical shell. This amounts to choosing the solution of
the form
\begin{equation}
x^\mu_i (s)=\{t(s),\, r(s),\,\theta_i\, ,\,\varphi_i\}\; ,
\quad \frac{d\theta_i}{ds}=0\; ,
\quad \frac{d\varphi_i}{ds}=0
\end{equation}
with $t(s)$ and $r(s)$ independent of $i$, and identifying $i$ with
the set $\{\theta_i\, ,\,\varphi_i\}$ . Then the motion of the source
boils down to a radial motion in an electric field of a single
particle with mass $M$ and charge $e$ . The electric field is the self 
field of the shell. An important fact is that the electric field
is discontinuous on a charged surface, and the force exerted by
the shell on itself is determined by one half of the sum of the
electric fields on both sides of the shell [8]. Writing the law
of motion of the shell in the form
\begin{equation}
r=\rho (t)
\end{equation}
one obtains
\begin{equation}
M\frac{d}{dt}\left(\frac{{\dot\rho}}{\sqrt{1-{\dot\rho}^2}}\right)
=e\,\frac{E_+ + E_-}{2}\Bigl |_{\textstyle\mbox{shell}}
\end{equation}
where $E_+$ and $E_-$ are the electric fields outside and inside
the shell.

Any spherically symmetric electromagnetic field is determined by
a single function $\e (t,r)$ which is the charge contained at the
time instant $t$ inside the sphere of area $4\pi r^2$ :
\begin{equation}
\e (t,r)=\bintegral \theta (r-{\bar r})\delta(t-{\bar t})
{\bar\nabla}_\mu {\bar t}\, j^\mu ({\bar x})\; .
\end{equation}
In terms of this function the solution of the conservation equation
$\nabla_\alpha j^\alpha =0$ is
\begin{equation}
-4\pi r^2 j^\mu =\Bigl(\nabla^\mu t\frac{\partial}{\partial r}
+\nabla^\mu r\frac{\partial}{\partial t}\Bigr)\e (t,r)\; ,
\end{equation}
and the solution of the Maxwell equations (2.8) is
\begin{equation}
F_{\mu\nu}=\left(\nabla_\mu r\nabla_\nu t -
\nabla_\mu t\nabla_\nu r\right)E
\end{equation}
with the electric field
\begin{equation}
E=\frac{\e (t,r)}{r^2}\; .
\end{equation}
The function $\e (t,r)$ must satisfy the condition of regularity
of the electric field at $r=0$
\begin{equation}
\e (t,0)=0
\end{equation}
and the normalization condition
\begin{equation}
\e (t,\infty)=e\; .
\end{equation}

Since $\ja$ in Eq. (2.10) is conserved, it is also of the form (3.5):
\begin{equation}
-4\pi r^2 \jm =\Bigl(\nabla^\mu t\frac{\partial}{\partial r}
+\nabla^\mu r\frac{\partial}{\partial t}\Bigr)\be (t,r)\; ,
\end{equation}
and, for $\be (t,r)$ , Eq. (2.10) yields the obvious result
\begin{equation}
\be (t,r)=e\,\theta (r-\rho(t))\; .
\end{equation}
Thus, owing to the conservation of the current $j^\alpha$ ,
which is a corollary of the expectation-value equations (2.11),
only one of these equations is independent. Finally, on account of
Eq. (3.7), the equation of motion of the shell (3.3) takes the form
\begin{equation}
M\frac{d}{dt}\left(\frac{{\dot\rho}}{\sqrt{1-{\dot\rho}^2}}\right)
=e\,\frac{\e_+ (t) + \e_- (t)}{2\rho^2}
\end{equation}
where
\begin{equation}
\e_\pm (t)=\e (t,\rho(t)\pm 0)\; .
\end{equation}
Eqs. (2.11), (3.5), and (3.10)-(3.13) make a closed set of equations
for $\e (t,r)$ and $\rho (t)$ .

The setting of the problem with the in-vacuum of quantum fields
implies that the external or mean fields generated by the source
are asymptotically static in the past [2]. Accordingly, it will be
assumed that, before some time instant $t=\t$ , the shell was kept
at some constant value of $r$, $r=\r$ , and next was let go.
Eq. (3.12) will thus be solved with the initial conditions
\begin{equation}
\rho\Bigl |_{\t} =\r\qquad ,\qquad {\dot\rho}\Bigl |_{\t} =0\; .
\end{equation}
The energy of this initial state is already affected by the static
vacuum polarization. However, for
\begin{equation}
m\r\stackrel{\displaystyle >}{\sim}1
\end{equation}
this effect is negligible (see Eq. (5.18) below), and it does not
make sense to consider $\r$ smaller than the Compton size of the
vacuum particle. Then, up to a small correction, the energy of
the shell (with the rest energy subtracted) retains its
classical value
\begin{equation}
\E =\frac{e^2}{2\r}\; ,
\end{equation}
and so does the acceleration of the shell at $t=\t$
\begin{equation}
{\ddot\rho}\Bigl |_{\t} =\frac{\E}{M}\,\frac{1}{\r}\; .
\end{equation}

Since the shell moves with acceleration, it creates particles from
the vacuum provided that its typical frequency exceeds the threshold 
of pair creation: $\hbar\nu >2mc^2$. At the high-frequency limit
$\hbar\nu\gg mc^2$ its vacuum radiation stops depending on the 
mass $m$ [4]. As seen from Eq. (3.17), the typical frequency $\nu$
is proportional to $\E/M$ :
\begin{equation}
\nu =\frac{\E}{M}\,\frac{1}{\r}\; .
\end{equation}
The bigger the ratio $\E/M$ , the bigger is the acceleration at $\t$ ,
and the more violent is the creation of particles. Therefore, it is
interesting to consider the ultrarelativistic shell $(\E/M)\gg 1$.
The latter condition can be enhanced to provide for the high-frequency
regime:
\begin{equation}
\frac{\E}{M}\gg m\r\; .
\end{equation}
At the same time, under condition (3.15) the shell does not probe the
small scales where the present description may break down. Assuming
both Eqs. (3.15) and (3.19) one switches over from the consideration
of the static vacuum polarization to studying the vacuum reaction
on a rapidly moving source creating pairs. This is the purpose of the
present work.

Without predetermining the law of motion $\rho (t)$ one may assume that,
beginning with $t=\t$ , the shell expands monotonically with an
increasing velocity ${\dot\rho}(t)$ which at $t=\infty$ reaches some
finite value ${\dot\rho}(\infty)$ . Then ${\dot\rho}(\infty)$ may
serve as a measure {\it at late time} of the acceleration at $\t$ .
The world line of the shell is shown in Fig. 1. As $(\E/M)\to\infty$ ,
the velocity ${\dot\rho}(t)$ approaches $1$ at all $t$ except in a
small sector near $t=\t$ . The world line of the shell approaches then
the broken line $N$ in Fig. 1. These assumptions are valid for the
classical motion of the shell
\begin{equation}
\frac{M}{\sqrt{1-{\dot\rho}^2}} + \frac{1}{2}\,\frac{e^2}{\rho}
=M+\E\; ,
\end{equation}
and they cannot be invalidated by the quantum corrections if the
coupling constant $\kappa^2$ is small.
}

\newpage
{\renewcommand{\theequation}{4.\arabic{equation}}

\begin{center}
\section{\bf    The retarded resolvent}
\end{center}

As in Refs. [3,4], it is convenient to express the resolvent in the
expectation-value equations through the operator
\begin{equation}
\H =\sqrt{\frac{2q}{\Box}} K_1(\sqrt{2q\Box})\; ,\qquad q<0
\end{equation}
depending on the parameter $q$, whose retarded kernel is of the form
\begin{equation}
\H X(x)=\frac{1}{4\pi}\cint \delta 
\left(\sigma (x,{\bar x})-q\right)X({\bar x})\; .
\end{equation}
Here $K_1$ is the order-1 Macdonald function, and $\sigma(x,{\bar x})$
is the world function: one half of the square of the geodetic distance
between the points $x$ and ${\bar x}$ [9]. The integration in Eq. (4.2)
is over the past sheet of the hyperboloid of equal geodetic distance
$\sqrt{-2q}$ from the observation point $x$.

The needed expression is provided by the formula
\begin{equation}
\frac{1}{\mu^2 -\Box}=\bes K_0 (\sqrt{2q\Box})
\end{equation}
involving the Macdonald and Bessel functions, and the result is the
following expression for the kernel of the retarded resolvent:
\begin{equation}
\frac{1}{\mu^2 -\Box}X=\bes \frac{d}{dq}\H X
\end{equation}
with $\H X$ in Eq. (4.2). If the test function is a tensor
\begin{equation}
X=X^{\mu_1\,\ldots\,\mu_n}\; ,
\end{equation}
Eq. (4.2) is an abbreviation of
\begin{equation}
\H X^{\mu_1\,\ldots\,\mu_n}(x)=\frac{1}{4\pi}\cint\delta
\left(\sigma(x,{\bar x})-q\right)
g^{\mu_1}_{\;{\bar\mu}_1}(x,{\bar x})\ldots
g^{\mu_n}_{\;{\bar\mu}_n}(x,{\bar x})
X^{{\bar\mu}_1\,\ldots\,{\bar\mu}_n}({\bar x})
\end{equation}
where $g^\mu_{\;{\bar\mu}}(x,{\bar x})$ is the propagator of the
geodetic parallel transport [9].

As an example of the use of Eq. (4.4) one may derive the retarded
kernel of the massless operator $1/\Box$ . For a test function
$X$ asymptotically static in the past, the operator $\H$ decreases
as $q\to -\infty$ [3]:
\begin{equation}
\H X\Bigl |_{q\to -\infty}\propto\frac{1}{\sqrt{-q}}\; .
\end{equation}
Thus one obtains
\begin{equation}
-\frac{1}{\Box}X(x)=\H X\Bigl |_{q=0}=\frac{1}{4\pi}\cint\delta
\left(\sigma(x,{\bar x})\right)X({\bar x})\; .
\end{equation}

When $X$ is a spherically symmetric scalar $X=X(t,r)$, and the
coordinates (2.3) are used, one has
\begin{equation}
\H X(x)=\frac{1}{2}\phint{\bar r}^2\int\limits^1_{-1}
d(\cos\omega)\,\delta(\sigma-q)X({\bar t},{\bar r})\; ,
\end{equation}
\begin{equation}
2\sigma=-(t-{\bar t})^2 +(r-{\bar r})^2 +2r{\bar r}(1-\cos\omega)
\end{equation}
where $\omega$ is the arc length between the points $(\theta ,\varphi )$
and $({\bar\theta},{\bar\varphi})$ on the unit two-sphere. Denote
${}^2\sigma$ the world function of the two-dimensional Lorentzian
section:
\begin{equation}
{}^2\sigma=-\frac{1}{2}(t-{\bar t})^2 +\frac{1}{2}(r-{\bar r})^2\; .
\end{equation}
It follows from Eq. (4.10) that, on the hyperboloid $\sigma=q$,
the range $-1<\cos\omega<1$ is equivalent to the following range
of variation of ${}^2\sigma$ :
\begin{equation}
q-2r{\bar r}<{}^2\sigma<q\; .
\end{equation}
Therefore, the result of the angle integration in Eq. (4.9) is
\begin{equation}
\H X(x)=\frac{1}{2r}\phint{\bar r}X({\bar t},{\bar r})
\theta(q-{}^2\sigma)\theta({}^2\sigma +2r{\bar r}-q)\; ,
\end{equation}
\begin{equation}
\frac{d}{dq}\H X(x)=\frac{1}{2r}\phint{\bar r}X({\bar t},{\bar r})
\left[\delta({}^2\sigma -q)-\delta({}^2\sigma +2r{\bar r} -q)\right]\; .
\end{equation}
In the past of the observation point $x$, the boundaries specified
by the $\theta$-functions in Eq. (4.13) are of the form
\begin{eqnarray}
{}^2\sigma -q=0\; :&\quad {\bar t}=t-\sqrt{(r-{\bar r})^2 -2q}\; ,\\
{}^2\sigma +2r{\bar r}-q=0\; :&\quad
{\bar t}=t-\sqrt{(r+{\bar r})^2 -2q}\; ,
\end{eqnarray}
and Eq. (4.13) can be rewritten as
\begin{equation}
\H X=\frac{1}{2r}\int\limits^\infty_0 d{\bar r}\,{\bar r}
\int\limits^{\textstyle t-\sqrt{(r-{\bar r})^2 
-2q}}_{\textstyle t-\sqrt{(r+{\bar r})^2 -2q}}
d{\bar t}\,X({\bar t},{\bar r})\; .
\end{equation}

Eq. (4.17) yields a simple expression in the case where the source
$X$ is static: $X(t,r)=X(r)$. In this case one obtains
\begin{equation}
\H X=\frac{1}{2r}\int\limits^\infty_0 d{\bar r}\,{\bar r}X({\bar r})
\left(\sqrt{(r+{\bar r})^2-2q} - \sqrt{(r-{\bar r})^2-2q}\right)
\end{equation}
and
\begin{equation}
\frac{1}{\mu^2-\Box}X=\frac{1}{2\mu r}\int\limits^\infty_0
d{\bar r}\,{\bar r}X({\bar r})\Bigl[\exp \Bigl(-\mu |r-{\bar r}|\Bigr)-
\exp \Bigl(-\mu (r+{\bar r})\Bigr)\Bigr]
\end{equation}
where use is made of the integral
\begin{equation}
\int\limits_{-\infty}^0 dq\,\frac{J_0 (\mu\sqrt{-2q})}{\sqrt{a^2-2q}}
=\frac{1}{\mu}\exp \Bigl(-\mu |a|\Bigr)\; .
\end{equation}
Of course, for a static $X$, expression (4.18) could be obtained
simpler by integrating in Eq. (4.9) first over ${\bar t}$
\begin{equation}
\H X=\frac{1}{2}\int\limits^\infty_0 d{\bar r}\,{\bar r}^2
\int\limits^1_{-1} d(\cos\omega)\,
\frac{X({\bar r})}{\sqrt{(r-{\bar r})^2 +2r{\bar r}(1-\cos\omega)-2q}}
\end{equation}
and next over $\cos\omega$ .
}

\newpage
{\renewcommand{\theequation}{5.\arabic{equation}}

\begin{center}
\section{\bf    The static vacuum polarization}
\end{center}

The propagator of parallel transport $g^\mu_{\;{\bar\mu}}(x,{\bar x})$
for the metric (2.3) projects on the basis vectors as follows:
\begin{eqnarray}
\nabla_\mu t\: g^\mu_{\; {\bar\mu}}(x,{\bar x})
&=&{\bar\nabla}_{\bar\mu}{\bar t}\; ,\\
\nabla_\mu r\: g^\mu_{\; {\bar\mu}}(x,{\bar x})
&=&\cos\omega\, {\bar\nabla}_{\bar\mu}{\bar r}+
{\bar r}\, {\bar\nabla}_{\bar\mu}\cos\omega\; .
\end{eqnarray}
One can choose any of the two projections to convert Eq. (2.11) into
a scalar equation. Specifically, one can use the fact that, owing
to Eq. (5.1), the operation of projecting on $\nabla t$ commutes with 
the action of any nonlocal form factor. Hence
\begin{equation}
\nabla_\alpha t\,j^\alpha =\nabla_\alpha t\,\ja
-\gamma (-\Box)\nabla_\alpha t\,\ja\; ,
\end{equation}
and by Eqs. (3.5) and (3.11)
\begin{equation}
\nabla_\alpha t\,j^\alpha =\frac{1}{4\pi r^2}\,
\frac{\partial}{\partial r}\,\e (t,r)\; ,
\end{equation}
\begin{equation}
\nabla_\alpha t\,\ja =\frac{e}{4\pi r^2}\,\delta 
\left(r-\rho (t)\right)\; .
\end{equation}

Strictly outside and inside the shell, the function (5.5) vanishes.
Therefore, in each of these regions, the local terms on the right-hand 
side of Eq. (5.3), i.e., the terms in which the current $\ja$ appears
at the observation point can be omitted. Specifically, the
subtraction term in the spectral formula (2.18) can be omitted.
As a result, one obtains the equation
\begin{equation}
\frac{1}{r^2}\,\frac{\partial}{\partial r}\,\e (t,r)=
-\frac{\kappa^2}{6}\sint\frac{1}{\mu^2 -\Box}\left(
\nabla_\alpha t\,\ja\right)
\end{equation}
which is valid separately in two regions for the point $(r,t)$ :
outside and inside the shell. Below, the notation $\varepsilon$
is used for the function
\begin{equation}
\varepsilon (t,r)=\theta\left(r-\rho (t)\right)-
\theta\left(\rho (t)-r\right)\; .
\end{equation}

The broken lines in Fig. 1 bound the future light cone of the point of
start. Denote $P$ (for {\it past}) the exterior of this cone. By 
causality, the region $P$ can be affected only by the static sector
of the evolution of the shell. Therefore, when calculating
$\e (t,r)$ for the point $(r,t)$ in $P$, the law of motion of the shell
can be taken $\rho (t)=\r$ . With this law, Eqs. (4.19) and (5.5) yield
straight away
\begin{eqnarray}
\frac{1}{\mu^2 -\Box}\left(\nabla_\alpha t\,\ja\right)&=&
\frac{e}{r}\,\frac{1}{8\pi\mu\,\r}\,\Bigl[
\exp\Bigl(-\mu |r-\r|\Bigr)-\exp\Bigl(-\mu (r+\r)\Bigr)\Bigr]\; ,\\
&&\hspace{75mm}{}(r,t)\in P\nonumber
\end{eqnarray}
and one obtains
\begin{eqnarray}
\frac{\partial}{\partial r}\,\e (t,r)&=&
-e\,\frac{r}{\r}\,\frac{\kappa^2}{24\pi}
\int\limits^\infty_{2m} d\mu\,\Gamma (\mu^2)\,
\Bigl[\exp\Bigl(-\mu |r-\r|\Bigr)-\exp\Bigl(-\mu (r+\r)\Bigr)\Bigr]\; ,
\hspace{5mm}\\
&&\hspace{97mm}{}(r,t)\in P\; .\nonumber
\end{eqnarray}
Here the expression in the square brackets is positive. Therefore,
in view of the condition $\Gamma (\mu^2 )\ge 0$, one has
\begin{equation}
\frac{1}{e}\,\frac{\partial}{\partial r}\,\e (t,r)<0
\end{equation}
both outside and inside the shell.

Eqs. (3.8) and (3.9) appear now in the role of boundary conditions
for the regions inside and outside the shell respectively, and
in both regions they fix the solution of Eq. (5.9). The solution
for $(r,t)\in P$ is
\begin{eqnarray}
\e (t,r)&=&e\,\frac{1+\varepsilon}{2}+ \frac{e}{\r}\,\frac{\kappa^2}{24\pi}
\int\limits^\infty_{2m}\frac{d\mu}{\mu^2}\,\Gamma(\mu^2)\\
&&{}\times\Bigl[(1+\varepsilon\mu r)\exp\Bigl(-\mu\varepsilon (r-\r)\Bigr)-
(1+\mu r)\exp\Bigl(-\mu (r+\r)\Bigr)\Bigr]\nonumber
\end{eqnarray}
and is, of course, static.

As shown below, a number of features of the static solution above
persists also beyond $P$, i.e., for a moving shell. These features
are summarized in Fig. 2. First of all, inside the shell there is
charge, and this charge is negative. This is a consequence of the
positivity of the spectral-mass function and the nonlocal nature
of the expectation-value equations. The retarded nonlocal form
factor collects the bare charge from the whole interior of the
past light cone of the observation point. Since the world line of
the shell crosses this cone at any location of the observation point,
the vacuum inside the shell gets polarized.

On the other hand, the total charge inside every sphere surrounding
the shell is positive. This occurs owing to a jump of $\e (t,r)$
across the shell, i.e., owing to the positive charge of the shell
itself. The jump is, however, infinite, and so are the values of
$\e (t,r)$ on both sides of the shell. This infinity, of different
signs inside and outside, develops in the Compton neighbourhood of
the shell. At a large distance from the shell, the polarization falls
off exponentially owing to the presence of the threshold $\mu\ge 2m$
in the spectral integral. Qualitatively, at each given $t$, the
$\e (t,r)$ for a moving shell has a similar shape.

The mechanism by which the singularity on the shell's surface
emerges is noteworthy since it appears repeatedly in the
consideration below. This mechanism is connected with the
convergence of the spectral integral in Eq. (5.11) at the
upper limit. The integrand in Eq. (5.11) provides an exponential
cut off at large spectral mass but only for $r\ne\r$. At
$r=\r$ , in view of the condition $\Gamma(\infty)=1$, the integral
becomes logarithmically divergent. This is none other than the
ultraviolet divergence of the charge renormalization. For an
observer at infinity, the shell appears as an electric monopole
screened by the vacuum. With $\e (t,\infty )$ normalized as in Eq. (3.9),
the unscreened monopole
\begin{equation}
\hspace{55mm}\e_+=\e (t,\r +0)\qquad\qquad\qquad (t\le\t )
\end{equation}
should be infinite.

However, in the present case the source is not a pointlike object.
The total charge inside the shell
\begin{equation}
\hspace{55mm}\e_-=\e (t,\r -0)\qquad\qquad\qquad (t\le\t )
\end{equation}
is also infinite and has the opposite sign. Owing to this fact,
the force moving the shell is finite. Indeed, with Eq. (5.11) one is
able to calculate the acceleration of the shell at $t=\t$
\begin{equation}
{\ddot\rho}\Bigl |_{\t}=\frac{e}{M\r^2}\,\frac{\e_+ +\e_-}{2}\; .
\end{equation}
Making the sum $\e_+ +\e_-$ in the spectral integral one obtains
unambiguously
\begin{equation}
\frac{\e_+ +\e_-}{2}=\frac{e}{2}+\frac{e}{\r}\,\frac{\kappa^2}{24\pi}
\int\limits^\infty_{2m}\frac{d\mu}{\mu^2}\,\Gamma (\mu^2)\,
\Bigl[1-(1+\mu\r )\exp (-2\mu\r)\Bigr]\; .
\end{equation}
This way of subtracting infinities is physically equivalent to giving
the shell a Compton width (see Section 9 for a refinement of this point).

The function in the square brackets in Eq. (5.15)
\begin{equation}
f(\mu\r )=1-(1+\mu\r )\exp (-2\mu\r )
\end{equation}
is positive since
\begin{equation}
\frac{d}{dx}f(x)>0\quad\mbox{for}\quad x>0
\end{equation}
and $f(0)=0$. Therefore, the force in Eq. (5.14) is in all cases 
repulsive. In the case $m\r\ll 1$ it even acquires an extra
amplifying factor $|\log m\r |$. However, under condition (3.15)
this force differs negligibly from its classical value:
\begin{equation}
\frac{\e_+ +\e_-}{2}=\frac{e}{2}+e\,\frac{\kappa^2}{24\pi}
\biggl[\frac{1}{m\r}\int\limits^\infty_2 \frac{dx}{x^2}\,
\Gamma (m^2x^2)+O\Bigl(\exp (-2m\r )\Bigr)\biggr]\; .
\end{equation}
Also, the charge inside the shell is then concentrated almost entirely
in the Compton neighbourhood of the shell, and so is the excess of
charge over $e$ outside the shell. In this way the correspondence
principle is fulfilled. On the other hand, no large scales or low 
energies can save one from the development of the singularity
within the Compton neighbourhood of the shell. Its appearance may 
be understood as a signal that a charge cannot be localized
more accurately than within a Compton neighbourhood. The charges
of the shell immersed in the real vacuum are always annihilated and 
created anew in a slightly different place. As a result, the shell
gets smeared to a Compton width. In this way the quantum uncertainty
manifests itself.

For the sake of comparison consider also a pointlike source. This is
the limiting case of the charged ball
\begin{eqnarray}
\be (t,r)=
\left\{
\begin{array}{ccc}
{\displaystyle e\,\frac{r^3}{r_0{}^3}}&,&
{\displaystyle \;\; r<r_0}\\
{\displaystyle e}&,&
{\displaystyle \;\; r>r_0}\\
\end{array}
\right.{}
\end{eqnarray}
as $r_0\to 0$. In this case, assuming that the observation point is
outside the ball, Eqs. (4.19) and (3.10) yield
\begin{equation}
\frac{1}{\mu^2 -\Box}\left(\nabla_\alpha t\,\ja\right)
=\frac{e}{4\pi r}\exp (-\mu r)\; .
\end{equation}
With the boundary condition (3.9) one then obtains from Eq. (5.6)
\begin{equation}
\e (t,r)=e+e\,\frac{\kappa^2}{12\pi}\int\limits^\infty_{2m}
\frac{d\mu}{\mu}\,\Gamma (\mu^2)\, (1+\mu r)\exp (-\mu r)\; .
\end{equation}
The electric field (3.7) with this $\e (t,r)$ can be written down as
\begin{equation}
E=-\frac{\partial}{\partial r}U\; ,
\end{equation}
\begin{equation}
U=\frac{e}{r}\biggl(1+\frac{\kappa^2}{12\pi}\int\limits^\infty_{2m}
\frac{d\mu}{\mu}\,\Gamma (\mu^2)\,\exp (-\mu r)\biggr)\; .
\end{equation}
With the spectral-mass function in Eq. (2.20) and $\kappa^2$ in
Eq. (2.21), this reproduces the textbook result for the "modified
Coulomb law". In fact, the Coulomb law is not modified as seen
from Eq. (3.7). What gets modified is the charge distribution.
The pointlike charge induces the same infinite screening as the
charged shell does.
}

\newpage
{\renewcommand{\theequation}{6.\arabic{equation}}

\begin{center}
\section{\bf    Solution for the moving shell}
\end{center}

Consider Eq. (4.13) in which $X(t,r)$ is identified with the charge
density (5.5). The lines (4.15) and (4.16) on the ${\bar r},{\bar t}$
plane bound the mapping on this plane of the past (sheet of the)
hyperboloid $\sigma (x,{\bar x})=q$ of the observation point $x$.
The observation point has the coordinates $r,t$ and is shown in
Fig. 3 along with the two boundaries of the mapping of its past
hyperboloid (the bold lines). At $q=0$ the past hyperboloid becomes
the past light cone whose mapping on the ${\bar r},{\bar t}$ plane
is bounded by the light lines in Fig. 3. The world line of the shell
${\bar r}=\rho ({\bar t})$ which is also shown in Fig. 3 crosses
the upper boundary of the past hyperboloid at some point
$r_+,t_+$ and the lower boundary at some point $r_-,t_-$.
These points are determined by the equations
\begin{eqnarray}
\left\{
\begin{array}{rcl}
{\displaystyle r_+}&{\displaystyle =}&
{\displaystyle \rho (t_+)\; ,}\\
{\displaystyle t_+}&{\displaystyle =}&
{\displaystyle t-\sqrt{(r-r_+)^2-2q}\; ,}\\
\end{array}
\right.{}
\end{eqnarray}
\begin{eqnarray}
\left\{
\begin{array}{rcl}
{\displaystyle r_-}&{\displaystyle =}&
{\displaystyle \rho (t_-)\; ,}\\
{\displaystyle t_-}&{\displaystyle =}&
{\displaystyle t-\sqrt{(r+r_-)^2-2q}\; ,}\\
\end{array}
\right.{}
\end{eqnarray}
and their locations on the world line of the shell depend on the location
of the observation point $r,t$ and on the value of $q$. At $q=0$ the
coordinates solving Eqs. (6.1) and (6.2) will be denoted
\begin{equation}
r_+^0\, ,\; t_+^0\, ,\; r_-^0\, ,\; t_-^0
\end{equation}
respectively.

With the notation above, Eqs. (4.13) and (5.5) yield
\begin{equation}
\H\left(\nabla_\alpha t\,\ja\right)=\frac{e}{8\pi r}
\int\limits^{\textstyle t_+}_{\textstyle t_-}
\frac{dt}{\rho (t)}\; .
\end{equation}
Using Eqs. (6.1), (6.2) one can calculate
\begin{equation}
\frac{\partial t_\pm}{\partial q}\equiv \frac{1}{A_\pm}\; ,
\end{equation}
\begin{equation}
A_+=(t-t_+)-(r-r_+){\dot\rho}(t_+)\; ,
\end{equation}
\begin{equation}
A_-=(t-t_-)+(r+r_-){\dot\rho}(t_-)\; ,
\end{equation}
and hence
\begin{equation}
\frac{d}{dq}\H\left(\nabla_\alpha t\,\ja\right)=
\frac{e}{8\pi r}\left(\frac{1}{r_+A_+}-\frac{1}{r_-A_-}\right)\; .
\end{equation}
With this expression, Eq. (4.4) yields
\begin{equation}
\frac{1}{\mu^2-\Box}\left(\nabla_\alpha t\,\ja\right)=
\frac{e}{8\pi r}\bes\left(\frac{1}{r_+A_+}-\frac{1}{r_-A_-}\right)\; ,
\end{equation}
and then from Eq. (5.6) one obtains
\begin{equation}
\frac{\partial}{\partial r}\,\e (t,r)=-r\,\frac{e\kappa^2}{48\pi}
\sint\bes\left(\frac{1}{r_+A_+}-\frac{1}{r_-A_-}\right)\; .
\end{equation}

Eq. (6.10) should now be integrated over $r$ along the line
$t=\mbox{const.}$ with the boundary conditions (3.8) and (3.9)
inside and outside the shell respectively. This integration
can be done explicitly, and the final result is
\begin{equation}
\e (t,r)=e\,\frac{1+\varepsilon}{2}+e\,\frac{\kappa^2}{24\pi}
\sint w(\mu,t,r)\; ,
\end{equation} 
\begin{equation}
w(\mu,t,r)=\frac{1}{\mu^2}\,\frac{1+\varepsilon}{2}+
\bes F(q,t,r)
\end{equation}
with $\varepsilon$ in Eq. (5.7), and
\begin{equation}
F(q,t,r)=\frac{1}{2}\biggl[\:
\int\limits^{\textstyle t_+}_{\textstyle t_-}\frac{dt}{\rho (t)}
+\log\Bigl((t-t_-)-(r+r_-)\Bigr)
-\log\Bigl((t-t_+)+(r-r_+)\Bigr)\biggr]\; .
\end{equation}
It follows from Eqs. (6.1), (6.2) that the arguments of both $\log$'s
in Eq. (6.13) are nonnegative.

For the proof of the result above, first use Eqs. (6.1), (6.2) to show
that the derivative of the function (6.13) is
\begin{equation}
\frac{\partial}{\partial r}\,F(q,t,r)=-\frac{r}{2}\:\left(
\frac{1}{r_+A_+}-\frac{1}{r_-A_-}\right)\; ,
\end{equation}
and thereby expression (6.11) satisfies the equation (6.10).
Next note that at $r=0$ the points $r_+,t_+$ and $r_-,t_-$ coincide.
Hence
\begin{equation}
F(q,t,0)=0\; ,
\end{equation}
and thereby expression (6.11) with $\varepsilon =-1$ satisfies the
boundary condition (3.8).

Finally, consider expression (6.11) with $\varepsilon =+1$ at the
limit where the observation point $r,t$ moves to spatial infinity:
$r\to\infty$ at a fixed $t$. At any $t$ and a sufficiently large
$r$, the point $r,t$ will enter the region $P$ which is affected 
only by the static sector of the evolution of the shell. Indeed,
as the observation point moves to spatial infinity, both points
$r_+,t_+$ and $r_-,t_-$ shift to the past along the world line
of the shell and turn out to be on its static sector. Therefore,
\begin{equation}
F(q,t,r)\Bigl |_{r\to\infty}=\F(q,r)\Bigl |_{r\to\infty}
\end{equation}
where
\begin{eqnarray}
\F(q,r)&=&\frac{1}{2}\biggl[\frac{1}{\r}\Bigl(\sqrt{(r+\r)^2-2q}
-\sqrt{(r-\r)^2-2q}\Bigr)\nonumber\\
&&{}+\log\Bigl(\sqrt{(r+\r)^2-2q}-(r+\r)\Bigr)\nonumber\\
&&{}-\log\Bigl(\sqrt{(r-\r)^2-2q}+(r-\r)\Bigr)\biggr]\; .
\end{eqnarray}
With this expression the integral in Eq. (6.12) can be calculated:
\begin{eqnarray}
\bes\F(q,r)=-\frac{1}{\mu^2}\,\frac{1+\varepsilon_0}{2}
+\frac{1}{2\mu^3\r}\hspace{51mm}\nonumber\\
\hspace{5mm}{}\times \Bigl[(1+\varepsilon_0\mu r)
\exp\Bigl(-\mu\varepsilon_0 (r-\r)\Bigr)
-(1+\mu r)\exp\Bigl(-\mu(r+\r)\Bigr)\Bigr]\; ,
\end{eqnarray}
\begin{equation}
\varepsilon_0 =\theta(r-\r)-\theta(\r-r)\; ,
\end{equation}
and it is seen why the explicit term in $1/\mu^2$ is introduced in
Eq. (6.12). Here use is made of the integrals
\begin{equation}
\bes\Bigl(\sqrt{a^2-2q}-\sqrt{-2q}\Bigr)=\frac{1}{\mu^3}
\Bigl[1-(1+\mu|a|)\exp(-\mu|a|)\Bigr]\; ,
\end{equation}
\begin{equation}
\bes\log\left(\frac{\sqrt{a^2-2q}\pm|a|}{\sqrt{-2q}}\right)=
\pm\frac{1}{\mu^2}\Bigl(1-\exp(-\mu|a|)\Bigr)\; ,
\end{equation}
and the result agrees with Eq. (5.11). Thus one obtains from Eqs.
(6.11) and (6.16)
\begin{eqnarray}
\e (t,r)\Bigl |_{r\to\infty}=e+\frac{e}{\r}\,\frac{\kappa^2}{24\pi}
\int\limits^\infty_{2m}\frac{d\mu}{\mu^2}\,\Gamma (\mu^2)
\Bigl(\exp(\mu\r)-\exp(-\mu\r)\Bigr)\hspace{25mm}\nonumber\\
{}\times (1+\mu r)\exp(-\mu r)
\Bigl|_{r\to\infty}
\end{eqnarray}
whence
\begin{equation}
\e (t,r)\Bigl |_{r\to\infty}=e+O\Bigl(\exp(-2mr)\Bigr)
\end{equation}
and thereby the boundary condition (3.9) is satisfied.
}

\newpage
{\renewcommand{\theequation}{7.\arabic{equation}}

\begin{center}
\section{\bf    Convergence of the spectral integral}
\end{center}

The integral (6.12) in $q$ always converges. Indeed, when the observation 
point $r,t$ is fixed, and $q\to -\infty$, the points $r_+,t_+$ and
$r_-,t_-$ again shift to the past and turn out to be on the static
sector of the world line of the shell. Therefore,
\begin{equation}
F(q,t,r)\Bigl |_{q\to -\infty}=\F(q,r)\Bigl |_{q\to -\infty}
\end{equation}
whence
\begin{equation}
F(q,t,r)\Bigl |_{q\to -\infty}=\frac{r\r^2}{(-2q)^{3/2}}\; .
\end{equation}
Since also
\begin{equation}
F(q,t,r)\Bigl |_{q\to 0}=O\Bigl(\log(-2q)\Bigr)
\end{equation}
as discussed below, the integral in Eq. (6.12) converges even at
$\mu=0$ and at any location of the observation point $r,t$.

The behaviour (7.3) is in all cases calculable directly from
Eqs. (6.13) and (6.1), (6.2) but its coefficient is different
for different locations of the observation point. For the
observation point outside the shell one obtains
\begin{equation}
F(q,t,r)\Bigl |_{r>\rho (t)}=\frac{1}{2}\log(-2q)+
\frac{1}{2}\sum^\infty_{n=0}a_n(t,r)(-2q)^n\; , \qquad q\to 0
\end{equation}
with
\begin{equation}
a_0(t,r)=-\log[4(r-r_+^0)(r+r_-^0)]+
\int\limits^{t_+^0}_{t_-^0}\frac{dt}{\rho (t)}
\end{equation}
whereas, for the observation point inside the shell,
\begin{equation}
F(q,t,r)\Bigl |_{r<\rho (t)}=\frac{1}{2}\sum^\infty_{n=0}
b_n(t,r)(-2q)^n\; , \qquad q\to 0
\end{equation}
with
\begin{equation}
b_0(t,r)=\log\frac{(r_+^0-r)}{(r_-^0+r)}+
\int\limits^{t_+^0}_{t_-^0}\frac{dt}{\rho (t)}\; .
\end{equation}
The result for the observation point on the shell is different
from both Eqs. (7.4) and (7.6):
\begin{equation}
F(q,t,r)\Bigl |_{r=\rho (t)}=\frac{1}{4}\log(-2q)+
\frac{1}{2}\sum^\infty_{n=0}c_n(t)(-2q)^{n/2}\; , \qquad q\to 0\; .
\end{equation}
Here the expansion is in half-integer powers with
\begin{equation}
c_0(t)=-\log2[\rho (t)+\rho (t_-^0)]+
\log\sqrt{\frac{1-{\dot\rho}(t)}{1+{\dot\rho}(t)}}+
\int\limits^{t}_{t_-^0}
\frac{d{\bar t}}{\rho({\bar t})}\; ,
\end{equation}
\begin{equation}
c_1(t)=\frac{1}{\sqrt{1-{\dot\rho}^2(t)}}\left(-\frac{1}{\rho(t)}+
\frac{1}{2}\:\frac{{\ddot\rho}(t)}{1-{\dot\rho}^2(t)}\right)\; ,
\end{equation}
and $t_-^0$ taken at $r=\rho(t)$. The cause of this discontinuity is 
in the fact that at $q=0$ the upper boundary of the hyperboloid
acquires a conic singularity (Fig. 3). When the observation point 
crosses the shell, the point $r_+^0,t_+^0$ passes through the
vertex of the cone.

The behaviour of $F(q,t,r)$ at $q\to 0$ determines the leading
(power) behaviour of the integral (6.12) at $\mu\to\infty$.
The latter behaviour can be obtained by integrating by parts
with the aid of the relation
\begin{equation}
J_0(\mu\sqrt{-2q})=\frac{2}{\mu^2}\,\frac{\partial}{\partial q}\,q
\,\frac{\partial}{\partial q}\,J_0(\mu\sqrt{-2q})\; .
\end{equation}
In the case (7.4) or (7.6) one may write
\begin{eqnarray}
\bes F(q,t,r)=\Bigl(q-\frac{2}{\mu^2}\,q\,\frac{\partial}{\partial q}
\Bigr)\sum^N_{n=o}\Bigl(\frac{2}{\mu^2}\Bigr)^n\Bigl(
\frac{\partial}{\partial q}\,q\,\frac{\partial}{\partial q}\Bigr)^n
\,F(q,t,r)\biggl |_{q=0}\nonumber\\
{}+\Bigl(\frac{2}{\mu^2}\Bigr)^{N+1}\bes\Bigl(
\frac{\partial}{\partial q}\,q\,\frac{\partial}{\partial q}
\Bigr)^{N+1}F(q,t,r)
\end{eqnarray}
where use is made of Eq. (7.2). Since, for any $N$, the integral on 
the right-hand side of Eq. (7.12) converges and decreases as
$\mu\to\infty$, one obtains
\begin{eqnarray}
\bes F(q,t,r)\Bigl |_{r>\rho (t)}&=&-\frac{1}{\mu^2}
+O\Bigl(\frac{1}{\mu^{2N}}\Bigr)\; ,\qquad \mu\to\infty\\
\bes F(q,t,r)\Bigl |_{r<\rho (t)}&=&
O\Bigl(\frac{1}{\mu^{2N}}\Bigr)\; ,\qquad\qquad\quad\:\mu\to\infty
\end{eqnarray}
where the remainder decreases faster than any power of $1/\mu^2$.
(It decreases exponentially, see below.)

In the case (7.8) the integration by parts as above can be done only 
once. One may write
\begin{equation}
\frac{\partial}{\partial q}\,q\,\frac{\partial}{\partial q}\,
F(q,t,\rho (t))=\frac{1}{\sqrt{-2q}}\Phi(\sqrt{-2q})
\end{equation}
where $\Phi(x)$ is analytic at $x=0$, and
\begin{eqnarray}
\bes\frac{\partial}{\partial q}\,q\,\frac{\partial}{\partial q}\,
F(q,t,\rho(t))=\frac{1}{\mu}\int\limits^\infty_0dx\,J_0(x)
\Phi\Bigl(\frac{x}{\mu}\Bigr)
=\frac{1}{\mu}\Bigl(\Phi(0)+{\cal O}\Bigr)\; ,\\
{\cal O}\to 0\; ,\; \mu\to\infty\nonumber
\end{eqnarray}
where
\begin{equation}
\Phi(0)=-\frac{1}{4}\,c_1(t)
\end{equation}
by Eq. (7.8). Hence one obtains
\begin{equation}
\bes F(q,t,r)\Bigl |_{r=\rho (t)}=-\frac{1}{2\mu^2}
-\frac{c_1(t)}{2\mu^3}+\frac{{\cal O}}{\mu^3}\; ,\qquad \mu\to\infty
\end{equation}
with $c_1(t)$ in Eq. (7.10).

As a result, in both regions outside and inside the shell, the 
behaviour of the function $w(\mu,t,r)$ in Eq. (6.12) is
\begin{equation}
w(\mu,t,r)\Bigl|_{r\ne\rho (t)}=O\Bigl(\frac{1}{\mu^{2N}}\Bigr)
\; ,\qquad \forall N\; ,\;\mu\to\infty
\end{equation}
whereas, on the shell,
\begin{equation}
w(\mu,t,\rho (t)\pm 0)=\pm\frac{1}{2\mu^2}+O\Bigl(\frac{1}{\mu^3}
\Bigr)\; ,\qquad \mu\to\infty\; .
\end{equation}
Recalling the condition $\Gamma(\infty)=1$, one infers that the
spectral-mass integral in Eq. (6.11) converges and defines the
function $\e (t,r)$ in all cases except in the case where the point
$r,t$ is on the world line of the shell. In the latter case the 
spectral integral diverges logarithmically, the divergent terms
having the same coefficients but different signs on the two sides
of the shell. It follows that the distribution $\e (t,r)$ is
singular on the shell's surface, and the next task is obtaining the
form of this singularity.
}

\newpage
{\renewcommand{\theequation}{8.\arabic{equation}}

\begin{center}
\section{\bf    Singularity of the electric field on the shell's
surface}
\end{center}

For obtaining the behaviour of $\e (t,r)$ on the shell's surface,
the behaviour of the function $w(\mu,t,r)$ at $\mu\to\infty$
should be known including the exponentially decreasing terms.
These are determined by the singularities of the function
$F(q,t,r)$ in the complex plane of the variable $z=\sqrt{-2q}$.

The function $F(q,t,r)$ will now be considered only {\it off}
the shell. It is convenient first to integrate by parts
\begin{equation}
\bes F(q,t,r)=-\frac{2}{\mu}\int\limits^\infty_0 d\sqrt{-2q}\,
J_1(\mu\sqrt{-2q})\,q\frac{\partial}{\partial q}F(q,t,r)
\end{equation}
and next use the analyticity of $q(\partial F/\partial q)$ at
$q=0$ to write
\begin{equation}
\int\limits^\infty_0 dx\,J_1(\mu x)\Bigl(
q\frac{\partial}{\partial q}F\Bigr)\biggl |_{2q=-x^2}=
\frac{1}{\mu}\Bigl(q\frac{\partial}{\partial q}F\Bigr)\biggl |_{q=0}
+\frac{1}{2}\mbox{ Re }\int\limits^\infty_{-\infty}dx\,
H_1^{(1)}(\mu x+{\rm i}\epsilon)\Bigl(q\frac{\partial}{\partial q}
F\Bigr)\biggl |_{2q=-x^2}
\end{equation}
where $H_1^{(1)}$ is the Hankel function. For the function
$w(\mu,t,r)$ both outside and inside the shell this gives
\begin{equation}
w(\mu,t,r)=-\frac{1}{\mu}\mbox{ Re }\int\limits_{\textstyle {\cal C}}dz\,
H_1^{(1)}(\mu z)\Bigl[q\frac{\partial}{\partial q}F(q,t,r)\Bigr]
\biggl |_{2q=-z^2}
\end{equation}
where the contour ${\cal C}$ passes above the real axis and closes
counter-clockwise in the upper half-plane. Here
\begin{eqnarray}
q\frac{\partial}{\partial q}F(q,t,r)\! &=&\!\frac{1}{4A_+}
\Bigl[\frac{2q}{r_+}-(r_+-r)\left(1-{\dot\rho}^2(t_+)\right)\Bigr]
-\frac{1}{4}{\dot\rho}(t_+)\nonumber\\
\! &-&\!\frac{1}{4A_-}\Bigl[\frac{2q}{r_-}-
(r_-+r)\left(1-{\dot\rho}^2(t_-)\right)\Bigr]+\frac{1}{4}{\dot\rho}(t_-)
\end{eqnarray}
with $A_\pm$ in Eqs. (6.6) and (6.7).

Of all singularities of the function (8.4) in the variable
$z=\sqrt{-2q}\:$, we are presently interested in the ones that have 
the least $|\mbox{Im }z|$ as the point $r,t$ approaches the shell.
These are easily identified with the solutions of the equation
$A_+=0$. Indeed, as $r\to\rho (t)$, these solutions shift to $q=0$
and, thereby, to $\mbox{Im }z=0$ whereas the remaining singularities 
stay at $\mbox{Im }z\ne 0$. This can be seen from the fact that,
apart from the factor $1/A_+$ , expression (8.4) with
$\mbox{Im }z=0$, i.e., with real $q\le 0$ is nonsingular including
at $r=\rho (t)$.

The equation
\begin{equation}
A_+=0
\end{equation}
along with Eq. (6.1) determines both $q$ and the point $r_+,t_+$.
Denote $q^*$ the solution for $q$, and $r^*,t^*$ the solution
for $r_+,t_+$. The solution for $r_+,t_+$ proves to be real.
Indeed, the point $r^*,t^*$ is defined by the equations
\begin{eqnarray}
t-t^*\! &=&\! (r-r^*){\dot\rho}(t^*)\; ,\\
r^*\! &=&\!\rho (t^*)
\end{eqnarray}
and is thus a point at which the world line of the shell crosses 
the line specified by Eq. (8.6). The latter line is shown in
Fig. 4 (line $L$). It passes through the observation point $r,t$
and, at least in some neighbourhood of this point, is {\it spacelike}.
This can be checked by calculating along $L$
\begin{equation}
\frac{dr^*}{dt^*}=\frac{1+(r-r^*){\ddot\rho}(t^*)}{{\dot\rho}(t^*)}\; .
\end{equation}
It follows that, at least when the observation point is sufficiently
close to the shell, the intersection at $r^*,t^*$ exists and is unique
(Fig. 4). The solution for $q$ is then real and {\it positive}:
\begin{equation}
2q^*=\eta^2\qquad ,\qquad \eta =|r-r^*|\sqrt{1-{\dot\rho}^2(t^*)}
\end{equation}
whence for $z$ one obtains two complex conjugate solutions
\begin{equation}
z=\pm{\rm i}\eta\; .
\end{equation}

Introducing a notation for the coefficient of $1/(4A_+)$ in Eq. (8.4),
one has
\begin{equation}
q\frac{\partial}{\partial q}F(q,t,r)\biggl|_{A_+\to 0}\, =
\frac{1}{4A_+}\left(\beta\Bigl |_{A_+=0}\right)\; ,
\end{equation}
and one may calculate
\begin{equation}
\frac{\partial A_+^2}{\partial q}=-2\alpha
\end{equation}
with
\begin{eqnarray}
\alpha &=&1-{\dot\rho}^2(t_+)-(r_+-r){\ddot\rho}(t_+)\; ,\\
\beta &=&\frac{2q}{r_+}-(r_+-r)\left(1-{\dot\rho}^2(t_+)\right)\; .
\end{eqnarray}
Both $\alpha$ and $\beta$ are finite and nonvanishing at $A_+=0$,
and, moreover,
\begin{equation}
\alpha\Bigl |_{A_+=0}\, >0
\end{equation}
at least when the observation point is sufficiently close to the shell.
It follows that the solutions (8.10) are branch points of the
function (8.4):
\begin{eqnarray}
A_+^2\Bigl |_{q\to q^*}&=&-2\alpha\Bigl |_{q=q^*}(q-q^*)+\ldots\nonumber\\
&=&\alpha\Bigl |_{q=q^*}(z^2+\eta^2)+\ldots\; ,
\end{eqnarray}
\begin{equation}
q\frac{\partial}{\partial q}F(q,t,r)\biggl |_{A_+\to 0}\, =
\left(\frac{\beta}{4\sqrt{\alpha}}\biggl |_{A_+=0}\right)
\frac{1}{\sqrt{z^2+\eta^2}}\; .
\end{equation}
Of the two branch points, the integral (8.3) picks up the one
in the upper half-plane: $z=+{\rm i}\eta$, and its contribution 
at large $\mu$ is
\begin{equation}
w(\mu,t,r)\to\frac{1}{\mu^2\eta}\left(\frac{\beta}{2\sqrt{\alpha}}
\biggl |_{A_+=0}\right)\exp(-\mu\eta)\; .
\end{equation}

The contribution (8.18) is the leading one as the observation point
$r,t$ approaches the shell. Summarizing the calculation above,
one obtains
\begin{eqnarray}
w(\mu,t,r)=\biggl[\frac{\varepsilon}{2\mu^2}\,\frac{r}{r^*}\,
\frac{\sqrt{1-{\dot\rho}^2(t^*)}}{\sqrt{1-{\dot\rho}^2(t^*)
-(r^*-r){\ddot\rho}(t^*)}}+O\Bigl(\frac{1}{\mu^3}\Bigr)\biggr]
\exp\left(-\mu|r^*-r|\sqrt{1-{\dot\rho}^2(t^*)}\right)\; ,\nonumber\\
r\to\rho (t)\; ,\;\mu\to\infty\hspace{1cm}
\end{eqnarray}
with $r^*,t^*$ the solution of the equations (8.6), (8.7). It follows
immediately from these equations that, when the observation point 
$r,t$ is on the shell, the point $r^*,t^*$ coincides with $r,t$
(see Fig. 4). Therefore, as $r\to\rho (t)$, one may expand
\begin{equation}
\rho (t^*)=\rho (t)+{\dot\rho}(t)(t^*-t)+O\left(\rho (t)-r\right)^2
\end{equation}
and in this way obtain the solution
\begin{eqnarray}
r^*-r=\frac{1}{1-{\dot\rho}^2(t)}\left(\rho (t)-r\right)+
O\left(\rho (t)-r\right)^2\; ,\\
t^*-t=\frac{{\dot\rho}(t)}{1-{\dot\rho}^2(t)}\left(\rho (t)-r\right)
+O\left(\rho (t)-r\right)^2\; ,\\
r\to\rho (t)\; .\nonumber
\end{eqnarray}
This brings Eq. (8.19) to its final form
\begin{equation}
w(\mu,t,r)=\biggl(\frac{\varepsilon}{2\mu^2}+O\Bigl(\frac{1}{\mu^3}
\Bigr)\biggr)\exp\biggl(-\mu\,\frac{|r-\rho (t)|}{\sqrt{1-
{\dot\rho}^2(t)}}\biggr)\; ,\qquad
r\to\rho (t)\; ,\;\mu\to\infty\; .
\end{equation}

It is seen from the latter expression that, with any law of motion
$\rho (t)$, the mechanism of formation of the singularity on the
shell's surface is one and the same. At $r=\rho (t)$, the integrand
in Eq. (6.11) loses the exponential cut off and becomes
$O(1/\mu^2)$, $\mu\to\infty$. The behaviour of $\e (t,r)$ as
$r\to\rho (t)$ can now be obtained by calculating the spectral-mass
integral (6.11) with the function (8.23):
\begin{equation}
\frac{\partial}{\partial\eta}\,\e (t,r)=-\frac{e\kappa^2}{24\pi}\,
\frac{\varepsilon}{\eta}\int\limits^\infty_{2m\eta}
dx\,\Gamma \Bigl(\frac{x^2}{\eta^2}\Bigr)\exp(-x)\; ,\qquad
\eta =\frac{|r-\rho (t)|}{\sqrt{1-{\dot\rho}^2(t)}}\to 0\; .
\end{equation}
The same result is obtained with Eq. (6.10). Since 
$\Gamma (\infty)=1$, one has
\begin{equation}
\e (t,r)\Bigl |_{r\to\rho (t)\pm 0}=\mp e\,\frac{\kappa^2}{24\pi}
\log|r-\rho (t)|\; .
\end{equation}
This behaviour is shown in Fig. 2. The electric field, as given 
in Eq. (3.7), differs from $\e (t,r)$ only by the factor $1/r^2$
which is finite and continuous across the shell.
}

\newpage
{\renewcommand{\theequation}{9.\arabic{equation}}

\begin{center}
\section{\bf    The force exerted by the charged shell on itself}
\end{center}

The singularity of the electric field on the shell's surface does
not affect the motion of the shell since it cancels in the sum
\begin{equation}
\e_+(t)+\e_-(t)=\e(t,\rho(t)+0)+\e(t,\rho(t)-0)
\end{equation}
which according to Eq. (3.12) determines the force exerted by the
shell on itself. Making the sum (9.1) in the spectral integral
(6.11) makes this cancellation unambiguous. Indeed, at $q<0$
the points $r_+,t_+$ and $r_-,t_-$ move along smooth trajectories
as the observation point crosses the shell (see Fig. 3). Therefore,
the function $F(q,t,r)$ with $q<0$ remains continuous and defines
\begin{equation}
{\cal F}(q,t)\equiv F(q,t,\rho(t))\; .
\end{equation}
The integral
\begin{equation}
\bes F(q,t,r)
\end{equation}
is also continuous. The function $w(\mu,t,r)$ is discontinuous
but finite and defines
\begin{equation}
2\,{\cal W}(\mu,t)\equiv w(\mu,t,\rho(t)+0)+w(\mu,t,\rho(t)-0)\; .
\end{equation}
Then from Eqs. (6.11), (6.12) one obtains
\begin{equation}
\e_+(t)+\e_-(t)=e+e\,\frac{\kappa^2}{12\pi}\sint{\cal W}(\mu,t)\; ,
\end{equation}
\begin{equation}
{\cal W}(\mu,t)=\frac{1}{2\mu^2}+\bes{\cal F}(q,t)\; ,
\end{equation}
and the function ${\cal F}(q,t)$ is given by expression (6.13)
with the insertion of $r=\rho(t)$.

The behaviour of the function ${\cal F}(q,t)$ at $q\to -\infty$
is determined by Eq. (7.2). The behaviour of this function at $q\to 0$
is obtained in Eq. (7.8), and the behaviour of the integral (9.6) 
with this function is obtained in Eq. (7.18). For ${\cal W}(\mu,t)$
this yields
\begin{equation}
{\cal W}(\mu,t)=O\Bigl(\frac{1}{\mu^3}\Bigr)\; ,\qquad\mu\to\infty\; .
\end{equation}
As a result, the spectral-mass integral (9.5) converges, and the
force exerted on the shell is finite. The effect of making the sum
(9.1) in the spectral integral is a cancellation of the $1/\mu^2$,
$\mu\to\infty$ terms (7.20) in the integrand.

Since the calculation above implies a subtraction of infinities,
it should be analysed what regularization does it correspond to
physically. The answer is contained in expression (8.23). Calculate 
$\e(t,r)$ for two close points $r_1,t_1$ and $r_2,t_2$ outside
and inside the shell respectively, and consider the sum
\begin{equation}
\e(t_1,r_1)+\e(t_2,r_2)\; .
\end{equation}
This is given by the spectral integral (6.11) with
\begin{equation}
w(\mu,t_1,r_1)+w(\mu,t_2,r_2)=\frac{1}{2\mu^2}
\left[\exp\left(-\mu\,\eta(t_1,r_1)\right)-
\exp\left(-\mu\,\eta(t_2,r_2)\right)\right]+
O\Bigl(\frac{1}{\mu^3}\Bigr)\; ,
\end{equation}
\begin{equation}
\eta(t,r)=\frac{|r-\rho(t)|}{\sqrt{1-{\dot\rho}^2(t)}}\; .
\end{equation}
The result (9.5) is recovered at the limit where the exponents in 
Eq. (9.9) tend to zero. Since the scale for $\mu$, set up by the
spectral integral, is $m$, the regularization implied in
$\e_+(t)$ and $\e_-(t)$ before their sum is made consists in a
fixation of the parameter
\begin{equation}
m\,\eta(t,r)=\mbox{const. }\ne 0\; .
\end{equation}
As soon as the sum (9.8) is made, the points $r_1,t_1$ and $r_2,t_2$
can be brought to the shell in any succession and along any pathes.
Since the $1/\mu^2$ term of expression (9.9) cancels in all cases,
the limit for the sum is finite and independent of the way the
regularization is removed.

The difference $|r-\rho(t)|$ that figures in expression (9.10) is the
proper distance from the point $r,t$ to the shell along the line
$t=\mbox{const}$ . But not this distance is made fixed in Eq. (9.11). 
The function $\eta(t,r)$ is the proper distance from the point
$r,t$ to the shell along the line {\it orthogonal to the world
line of the shell}. Indeed, this function was introduced in Eq. (8.9),
and originally it had the form
\begin{equation}
\eta(t,r)=\sqrt{(r-r^*)^2-(t-t^*)^2}
\end{equation}
where $r^*,t^*$ is the point on the world line of the shell connected 
with the point $r,t$ by the line $L$ (Fig. 4). It is easy to check
from Eq. (8.8) that, up to $O(r-r^*)$, the line $L$ is orthogonal
to the world line of the shell at the point of their intersection.

The final inference is that the subtraction of infinities in the
spectral integral is physically equivalent to giving the shell
a Compton width in the direction orthogonal to its world line.
The lines on the $r,t$ plane specified by Eq. (9.11):
\begin{equation}
\frac{|r-\rho(t)|}{\sqrt{1-{\dot\rho}^2(t)}}=
\frac{\mbox{const.}}{m}
\end{equation}
mark the band of quantum uncertainty around the world line of
the shell. This band is shown in Fig. 5, and it narrows as the
speed of expansion increases.
}

\newpage
{\renewcommand{\theequation}{10.\arabic{equation}}

\begin{center}
\section{\bf    The ultrarelativistic limit}
\end{center}

It will now be shown that the force exerted by the shell on itself
is singular at the ultrarelativistic limit. This is the limit at
which the world line of the shell approaches the world line of
an outgoing light ray (the line $N$ in Fig. 1). To be more precise,
we consider a family of functions $\rho(t)$, for which
\begin{equation}
1-{\dot\rho}(t)=\delta\,f(\delta,t)\; ,\qquad\delta\to 0\; .
\end{equation}
Here $\delta$ is a parameter (function of the initial data), and
it is assumed that, at all $t>\t\:$, $f(\delta,t)$ has a finite
limit as $\delta\to 0$, whereas, at $t=\t\:$,
\begin{equation}
f(\delta,\t)=\frac{1}{\delta}\; .
\end{equation}
The function $f$ can be normalized as $f(\delta,\infty)=1$,
and then
\begin{equation}
\delta=1-{\dot\rho}(\infty)\; .
\end{equation}
Eq. (10.1) generalizes the form that the classical law of motion has 
as $(M/\E)\to 0$. Indeed, with $\E$ and $\r$ taken for independent
data, Eq. (3.20) can be written as
\begin{equation}
\frac{1}{\sqrt{1-{\dot\rho}^2}}=\frac{\E}{M}
\Bigl(1-\frac{\r}{\rho}\Bigr)+1\; .
\end{equation}
However, Eq. (10.1) does not predetermine the dependence of the velocity
on energy. The limiting form of $\rho(t)$ at $\delta=0$ is
\begin{eqnarray}
\rol(t)=\rho(t)\biggl|_{\delta=0}=
\left\{
\begin{array}{lc}
{\displaystyle \r\; ,}&{\displaystyle\qquad t<\t\; ,}\\
{\displaystyle \r+(t-\t)\; ,}&{\displaystyle\qquad t>\t\; .}\\
\end{array}
\right.{}
\end{eqnarray}
This world line is shown in Fig. 6.

Consider the function (9.2) for the shell obeying the law of motion
(10.1). The line $N$ in Figs. 1 and 6 is the boundary of the region 
$P$ considered in Section 5. Therefore, when the point $r,t$ in the
argument of the function $F(q,t,r)$ is on the line $N$, the
respective points $r_+,t_+$ and $r_-,t_-$ are, at all $q$,
at the static sector of the world line of the shell. For the function
(9.2) with $\rho(t)$ in Eq. (10.5) this yields the result
\begin{equation}
{\cal F}(q,t)\biggl|_{\delta=0}=\F(q,\rol(t))
\end{equation}
where the function $\F(q,r)$ is given in Eq. (6.17). The integral
that figures in Eq. (9.6) is then already calculated in Eq. (6.18).
One obtains
\begin{equation}
{\cal W}(\mu,t)\biggl|_{\delta=0}=-\frac{1}{2\mu^2}+
\frac{(1+\mu\rol(t))}{2\mu^3\r}\Bigl\{
\exp\Bigl[-\mu\Bigl(\rol(t)-\r\Bigr)\Bigr]-
\exp\Bigl[-\mu\Bigl(\rol(t)+\r\Bigr)\Bigr]\Bigr\}\; .
\end{equation}
For $t\le\t$ this brings one back to Eq. (5.15), but for $t>\t$ one has
\begin{equation}
{\cal W}(\mu,t)\biggl|_{\delta=0}=-\frac{1}{2\mu^2}\Bigl(1+
{\cal O}\Bigr)\; ,\qquad {\cal O}\to 0\; ,\;\mu\to\infty
\end{equation}
and the integral in Eq. (9.5) diverges at large $\mu$ :
\begin{equation}
[\e_+(t)+\e_-(t)]\biggl|_{\delta=0}\, =\infty\; ,\qquad t>\t\; .
\end{equation}
The force exerted on the shell is infinite at the ultrarelativistic
limit.

The null limit (10.5) for $\rho(t)$ is never reached exactly even
with the classical motion, and the next task is obtaining the
asymptotic behaviour of the force as $\rho(t)$ approaches the
null limit. For that consider any point with a given $t>\t$ on
a timelike world line of the shell. When this point is sufficiently
close to the line $N$, the respective point $r_-,t_-$ is, at all
$q$, at the static sector of the evolution of the shell (see Fig.3).
However, for the point $r_+,t_+$ this is not the case. Rather the
range of variation of $q$ should be divided into two: the one
for which $t_+<\t$ and the one for which $t_+>\t\:$. It follows
from Eq. (6.1) that the former range is
\begin{equation}
-\infty<q<-\frac{1}{2}s^2(t)\; ,
\end{equation}
and the latter is
\begin{equation}
-\frac{1}{2}s^2(t)<q<0
\end{equation}
where
\begin{equation}
s(t)=\sqrt{(t-\t)^2-(\rho(t)-\r)^2}\; .
\end{equation}
Only in the range (10.10) does one have
\begin{equation}
{\cal F}(q,t)=\F(q,\rho(t))\; .
\end{equation}

Expression (9.6) first integrated by parts as in Eq. (8.1):
\begin{equation}
{\cal W}(\mu,t)=\frac{1}{2\mu^2}-\frac{2}{\mu}\int\limits^\infty_0
d\sqrt{-2q}\,J_1(\mu\sqrt{-2q})\, q\frac{\partial}{\partial q}
{\cal F}(q,t)
\end{equation}
may now be written in the form
\begin{eqnarray}
{\cal W}(\mu,t)=\frac{1}{2\mu^2}\!
&-&\!\frac{2}{\mu}\int\limits^\infty_0 d\sqrt{-2q}\,J_1(\mu\sqrt{-2q})
\, q\frac{\partial}{\partial q}\F(q,\rho(t))\nonumber\\
{}\!&+&\!\frac{2}{\mu}\int\limits^{\textstyle s(t)}_0 
d\sqrt{-2q}\,J_1(\mu\sqrt{-2q})
\, q\frac{\partial}{\partial q}\F(q,\rho(t))\nonumber\\
{}\!&-&\!\frac{2}{\mu}\int\limits^{\textstyle s(t)}_0 
d\sqrt{-2q}\,J_1(\mu\sqrt{-2q})
\, q\frac{\partial}{\partial q}{\cal F}(q,t)\; .
\end{eqnarray}
It will be noted that $s(t)$ is the two-dimensional geodetic distance
between the point of start and a point on the shell. Therefore,
when the latter point is on the null line $N$, $s(t)$ vanishes.
Indeed, the insertion of the limiting form (10.5) for $\rho(t)$
in Eq. (10.12) yields
\begin{equation}
s(t)\biggl|_{\delta=0}=0\; ,\qquad t>\t\; .
\end{equation}
With $s(t)=0$, the last two integrals in Eq. (10.15) vanish, and one 
recovers the result (10.7). Thus $s(t)$ serves in Eq. (10.15)
as a parameter of proximity of the law $\rho(t)$ to its ultrarelativistic
limit.

For obtaining the asymptotic behaviour of ${\cal W}(\mu,t)$
as $s(t)\to 0$, rewrite the last two integrals in Eq. (10.15) as
\begin{eqnarray}
{}\!&{}&\!\frac{2}{\mu}\, s(t)\int\limits^1_0 dx\,J_1(x\mu s(t))\,
\biggl[q\frac{\partial}{\partial q}\F(q,\rho(t))\biggr]
\biggl |_{2q=-x^2s^2(t)}\nonumber\\
{}\!&-&\!\frac{2}{\mu}\, s(t)\int\limits^1_0 dx\,J_1(x\mu s(t))\,
\biggl[q\frac{\partial}{\partial q}{\cal F}(q,t)\biggr]
\biggl |_{2q=-x^2s^2(t)}
\end{eqnarray}
and recall that ${\cal W}(\mu,t)$ is needed at large $\mu$. The
approximation of interest is, therefore,
\begin{equation}
s(t)\to 0\; ,\quad \mu s(t)=\mbox{finite}\; .
\end{equation}
At this limit, the behaviours of the integrals (10.17) are obtained
by expanding
\begin{eqnarray}
\biggl[q\frac{\partial}{\partial q}\F(q,\rho(t))\biggr]
\biggl |_{q\to 0}&=&\frac{1}{2}+O(q)\; ,\qquad t>\t\\
\biggl[q\frac{\partial}{\partial q}{\cal F}(q,t)\biggr]
\biggl |_{q\to 0}&=&\frac{1}{4}+O(\sqrt{-2q})\; .
\end{eqnarray}
Here use is made of Eqs. (7.4) and (7.8). Using also the explicit form
(10.7) of the first two terms in Eq. (10.15), one obtains finally
\begin{equation}
{\cal W}(\mu,t)=-\frac{J_0(\mu s(t))}{2\mu^2}\Bigl(1+{\cal O}\Bigr)
\; ,\qquad {\cal O}\to 0\; ,\;\mu\to\infty\; ,\; s(t)\to 0\; .
\end{equation}

Eq. (10.21) is the sought for asymptotic formula for the ultrarelativistic
motion. Setting in it $s(t)=0$, one recovers the limiting behaviour (10.8)
which caused the divergence of the integral in Eq. (9.5). With the function
(10.21) this integral converges:
\begin{equation}
[\e_+(t)+\e_-(t)]\biggl|_{s(t)\to 0}=e-e\,\frac{\kappa^2}{24\pi}
\int\limits^\infty_{4m^2}\frac{d\mu^2}{\mu^2}\,\Gamma (\mu^2)\,
J_0(\mu s(t))\; ,
\end{equation}
and its behaviour as $s(t)\to 0$ can be found by calculating
\begin{equation}
\frac{\partial}{\partial s}\int\limits^\infty_{4m^2}
\frac{d\mu^2}{\mu^2}\,\Gamma (\mu^2)\,J_0(\mu s )=
-\frac{2}{s}\int\limits^\infty_{2ms} dx\,\Gamma\Bigl(\frac{x^2}{s^2}
\Bigr)\, J_1(x)\; .
\end{equation}
Since $\Gamma (\infty)=1$, one obtains
\begin{equation}
\e_+(t)+\e_-(t)=e+e\,\frac{\kappa^2}{12\pi}\log(ms(t))
+\kappa^2O(1)
\end{equation}
where $O(1)$ denotes the terms that remain finite at the 
ultrarelativistic limit.

The $s(t)$ in Eq. (10.12) can be represented in the form
\begin{equation}
s(t)=(t-\t)\sqrt{1-{\dot\rho}^2({\tilde t})}\; ,\qquad t>\t
\end{equation}
where ${\tilde t}$ is some time instant between $\t$ and $t$.
By Eq. (10.1),
\begin{equation}
\log\left(1-{\dot\rho}^2({\tilde t})\right)=
\log\left(1-{\dot\rho}^2(t)\right)+O(1)\; ,
\end{equation}
and, therefore, expression (10.24) may finally be written as
\begin{equation}
\e_+(t)+\e_-(t)=e+e\,\frac{\kappa^2}{24\pi}
\log\left(1-{\dot\rho}^2(t)\right)+\kappa^2O(1)\; .
\end{equation}

By derivation, the remainder $O(1)$ in Eq. (10.27) is bounded
uniformly in energy but not necessarily in time. Because of
Eq. (10.2), one may worry about the vicinity of $t=\t\:$.
However, also at $t=\t\:$, expression (10.27) yields the correct
result, Eq. (5.18), provided that condition (3.15) is fulfilled.
Since $\kappa^2$ is small, all bounded terms of order $\kappa^2$
may be regarded as negligible corrections. The term $\kappa^2O(1)$
in Eq. (10.27) can then be discarded for all times and energies.
}

\newpage
{\renewcommand{\theequation}{11.\arabic{equation}}

\begin{center}
\section{\bf    Vacuum back-reaction on the motion of the shell}
\end{center}

The expression (10.27) with the term $\kappa^2O(1)$ discarded is to
be inserted in Eq. (3.12). Then the equation of motion of the shell
closes and takes the form
\begin{equation}
M\frac{d}{dt}\left(\frac{{\dot\rho}}{\sqrt{1-{\dot\rho}^2}}\right)=
\frac{e^2}{2\rho^2}\Bigl[1+\frac{\kappa^2}{24\pi}
\log(1-{\dot\rho}^2)\Bigr]\; .
\end{equation}
The last term on the right-hand side of this equation is the force
of the vacuum reaction. As will be clear in the next section, this
is the force of the back-reaction of a radiation produced by the charged
shell in the vacuum.

The force of the vacuum
back-reaction depends on the velocity. Nevertheless, the equation of
motion (11.1) admits the energy integral:
\begin{equation}
M\int\limits_{\textstyle 1}^{\textstyle 1/\sqrt{1-{\dot\rho}^2}}
\frac{dx}{\displaystyle 1-\frac{\kappa^2}{12\pi}\log x}+
\frac{1}{2}\,\frac{e^2}{\rho}=\E
\end{equation}
which at $\kappa^2 =0$ goes over into the classical law (3.20). 
That the constant $\E$ is indeed the energy of the initial state
is seen from the fact that, at ${\dot\rho}=0$, one recovers
Eq. (3.16).

There is
no problem with the singularity of the integral in Eq. (11.2). It is never
reached. As in Eq. (3.20), for a given energy, the velocity ${\dot\rho}$
reaches its maximum value at $\rho =\infty$ but the value is now
different:
\begin{equation}
\int\limits_{\textstyle 1}^{\textstyle 1/\sqrt{1-{\dot\rho}^2 (\infty)}}
\frac{dx}{\displaystyle 1-\frac{\kappa^2}{12\pi}\log x}=\frac{\E}{M}\; .
\end{equation}
As in Eq. (3.20), ${\dot\rho}(\infty)$ grows with $\E/M$ {\it but not
up to} {1}:
\begin{equation}
{\dot\rho}(\infty)=1-\frac{1}{2}\exp\left(-\frac{24\pi}{\kappa^2}\right)
\; ,\qquad \frac{\E}{M}\to\infty
\end{equation}
and this is the principal consequence of the vacuum back-reaction.

Eq. (11.2) is surprising. The coupling to the vacuum charges does not
change the electric potential\footnote{Under condition (3.15).} as one
could expect. It changes {\it the kinematics of motion} as relativity
theory does. Furthermore, within a given type of coupling, this
change is universal. It does not depend on the parameters of the
source, only on the coupling constant $\kappa^2$. There emerges
a new kinematic bound on the velocity of a charged body. As shown
below, this bound is crucial for the maintenance of the
conservation laws.
}

\newpage
{\renewcommand{\theequation}{12.\arabic{equation}}

\begin{center}
\section{\bf    Emission of charge}
\end{center}

The singularity of $\e (t,r)$ on the shell's surface, as calculated
in Section 8, has an important feature. Namely, the coefficient
of the divergent $\log$ in Eq. (8.25), or, equivalently, the coefficient
of $1/\mu^2$ in Eq. (7.20) is constant in time. This suggests that
the singularity comes from the static contribution which is
present in $\e_\pm(t)$ but cancels in the difference
\begin{equation}
\e_\pm(t_1)-\e_\pm(t_2)\; .
\end{equation}
The flux of charge across the shell should, therefore, be finite.

Indeed, from Eqs. (6.11), (6.12), and (9.2) one obtains
\begin{equation}
\e_\pm(t_1)-\e_\pm(t_2)=e\,\frac{\kappa^2}{24\pi}\sint
\Bigl[w(\mu,t_1,\rho(t_1)\pm 0)-
w(\mu,t_2,\rho(t_2)\pm 0)\Bigr]\; ,
\end{equation}
\begin{equation}
w(\mu,t_1,\rho(t_1)\pm 0)-
w(\mu,t_2,\rho(t_2)\pm 0)=\bes\,\Bigl[{\cal F}(q,t_1)-
{\cal F}(q,t_2)\Bigr]\; .
\end{equation}
It follows that, first, the quantity (12.3) is continuous across
the shell:
\begin{eqnarray}
&{}&w(\mu,t_1,\rho(t_1)+0)-
w(\mu,t_2,\rho(t_2)+0)\nonumber\\
&=&w(\mu,t_1,\rho(t_1)-0)-
w(\mu,t_2,\rho(t_2)-0)\; ,
\end{eqnarray}
and, therefore,
\begin{equation}
\e_+(t_1)-\e_+(t_2)=\e_-(t_1)-\e_-(t_2)\; .
\end{equation}
Second, by Eq. (7.18) the quantity (12.3) is $O(1/\mu^3)$,
$\mu\to\infty$, and, therefore, the difference (12.2) is finite.

For obtaining the flux of charge across the shell, no new calculation 
is needed. Denote
\begin{eqnarray}
\Delta e&=&\e_+(-\infty)-\e_+(\infty)\nonumber\\
&=&\e_-(-\infty)-\e_-(\infty)\; .
\end{eqnarray}
This is the charge emitted by the shell for the whole of its history.
Owing to Eq. (12.5), one may write
\begin{eqnarray}
\Delta e&=&\frac{1}{2}\left[\e_+(-\infty)+\e_-(-\infty)\right]\nonumber\\
&&{}-\frac{1}{2}\left[\e_+(\infty)+\e_-(\infty)\right]
\end{eqnarray}
and thereby relate the radiation of charge to the force of its
back-reaction. The latter has already been considered in
Sections 9-11. For the ultrarelativistic shell one has from Eq. (10.27)
\begin{equation}
\e_+(\infty)+\e_-(\infty)=e+e\,\frac{\kappa^2}{24\pi}
\log\left(1-{\dot\rho}^2(\infty)\right)+\kappa^2O(1)\; ,
\end{equation}
and, from Eq. (5.18),
\begin{equation}
\e_+(-\infty)+\e_-(-\infty)=e+\kappa^2O\left(\frac{1}{m\r}\right)\; .
\end{equation}
Hence
\begin{equation}
\Delta e=-e\,\frac{\kappa^2}{48\pi}\log\left(1-{\dot\rho}^2
(\infty)\right)+\kappa^2O(1)\; .
\end{equation}

Also the instantaneous radiation flux can readily be estimated.
Let $t_1<t_2$ be two time instants belonging to the epoch of the
rapid expansion of the shell. The amount of charge emitted by
the shell for the time between $t_1$ and $t_2$ is the quantity (12.1):
\begin{equation}
\e_\pm(t_1)-\e_\pm(t_2)=e\,\frac{\kappa^2}{48\pi}
\log\frac{1-{\dot\rho}^2(t_1)}{1-{\dot\rho}^2(t_2)}
+\kappa^2O(1)\; .
\end{equation}
From Eq. (10.1) one infers that this is a negligible amount:
\begin{equation}
\e_\pm(t_1)-\e_\pm(t_2)=\kappa^2O(1)\; ,\qquad \t<t_1<t_2\; .
\end{equation}
However, if in Eq. (12.11) one takes $\t=t_1\,$, i.e., if one
calculates the amount of charge emitted from the
beginning of expansion by the instant $t$, the result will be different:
\begin{eqnarray}
\e_\pm(\t)-\e_\pm(t)&=&-e\,\frac{\kappa^2}{48\pi}
\log\left(1-{\dot\rho}^2(t)\right)+\kappa^2O(1)\nonumber\\
&=&-e\,\frac{\kappa^2}{48\pi}
\log\left(1-{\dot\rho}^2(\infty)\right)+\kappa^2O(1)\; ,
\qquad \t<t\; .
\end{eqnarray}
This is easy to understand. The cause of the vacuum particle
creation is the acceleration of the source. The shell radiates
at a short stage of its evolution near $t=\t$ where its
acceleration is maximum [3]. Almost all the emitted charge
$\Delta e$ is released at this stage. Therefore, up to a small
correction, the quantity (12.13) is constant.

Thus the rate of emission of charge by the ultrarelativistic shell is
\begin{equation}
\frac{\Delta e}{e}=-\frac{\kappa^2}{48\pi}
\log\left(1-{\dot\rho}^2(\infty)\right)\; ,\qquad
{\dot\rho}(\infty)\to 1\; .
\end{equation}
The rate of emission of energy was calculated in Ref. [3]. Generalized
properly, this calculation yields the same result as in Eq. (12.14):
\begin{equation}
\frac{\Delta\E}{\E}=-\frac{\kappa^2}{48\pi}
\log\left(1-{\dot\rho}^2(\infty)\right)\; ,\qquad
{\dot\rho}(\infty)\to 1\; .
\end{equation}
One sees that the radiation rate grows unboundedly as the motion
of the shell approaches the ultrarelativistic limit. Inserting
in Eqs. (12.14) and (12.15) the ${\dot\rho}(\infty)$ calculated
from the {\it classical} law of motion (3.20), one obtains the
result [3]:
\begin{equation}
\frac{\Delta e}{e}=\frac{\Delta\E}{\E}=\frac{\kappa^2}{24\pi}
\log\frac{\E}{M}\; ,\qquad\frac{\E}{M}\to\infty
\end{equation}
which manifestly contradicts the conservation laws. However,
the result (12.16) does not take into account the back-reaction
of radiation. The vacuum friction does not allow the velocity
of the source to approach the speed of light closer than the
limit (11.4). The insertion of the expression (11.4) in 
Eqs. (12.14) and (12.15) restores the conservation laws:
\begin{equation}
\frac{\Delta e}{e}=\frac{\Delta\E}{\E}=\frac{1}{2}\; ,\qquad
\frac{\E}{M}\to\infty\; .
\end{equation}
Up to 50\% of energy and charge can be extracted from the source
by raising its initial energy.
}

\newpage

\begin{center}
\section*{\bf Acknowledgments}
\end{center}

The appearance of this article owes to a grant of the Italian
Ministry for Foreign Affairs and Landau Network - Centro Volta.
G.V. is supported also by the Russian Foundation for Fundamental
Research (Grant No. 99-02-18107).

\newpage

\begin{center}
\section*{\bf References}
\end{center}

\begin{enumerate}
\item G. A. Vilkovisky, Phys. Rev. Lett. 83 (1999) 2297
[hep-th/9906241].
\item B. S. DeWitt, Phys. Rep. C 19 (1975) 295.
\item A. G. Mirzabekian and G. A. Vilkovisky, Phys. Lett. B 414
(1997) 123; Ann. Phys. 270 (1998) 391 [gr-qc/9803006].
\item G. A. Vilkovisky, Phys. Rev. D 60 (1999) 065012
[hep-th/9812233].
\item G. A. Vilkovisky, Class. Quantum Grav. 9 (1992) 895.
\item A. O. Barvinsky, Yu. V. Gusev, G. A. Vilkovisky, and
V. V. Zhytnikov, J. Math. Phys. 35 (1994) 3525.
\item A. O. Barvinsky and G. A. Vilkovisky, Nucl. Phys. B 282
(1987) 163.
\item I. E. Tamm, {\it Foundations of Electricity Theory}
(Gostekhizdat, Moscow, 1956).
\item B. S. DeWitt, {\it Dynamical Theory of Groups and Fields}
(Gordon and Breach, New York, 1965).
\end{enumerate}

\newpage

\begin{center}
\section*{\bf Figure captions}
\end{center}

\begin{itemize}
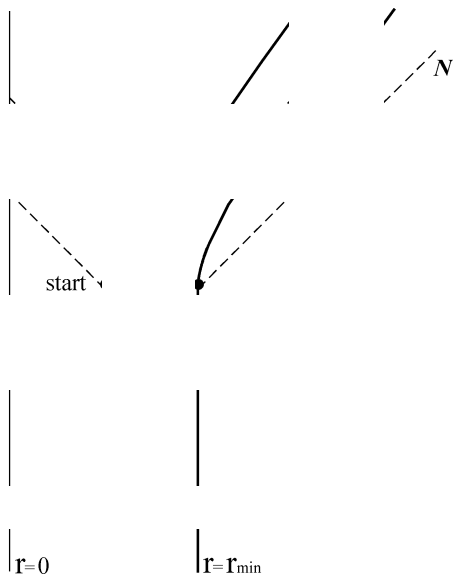
\item[Fig.1.] The world line of the shell on the $r,t$ plane.
The broken lines bound the future light cone of the point of
start. The broken line $N$ is the world line of the outgoing
radial light ray.
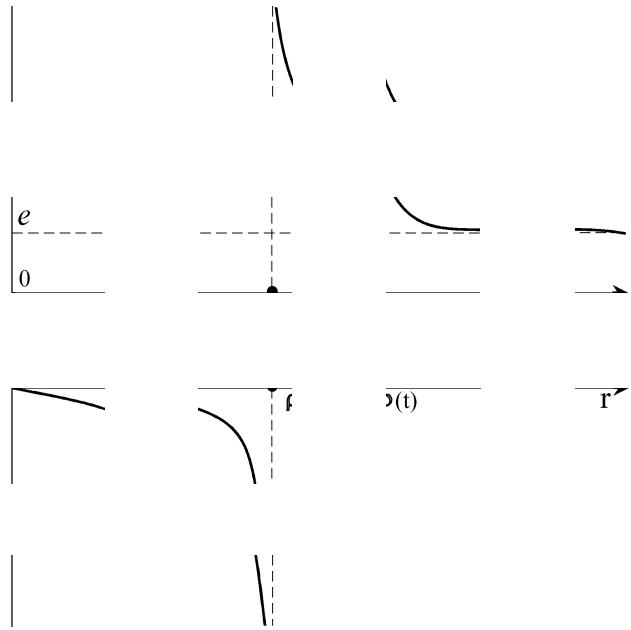
\item[Fig.2.] The function $\e (t,r)$ for a given $t$.
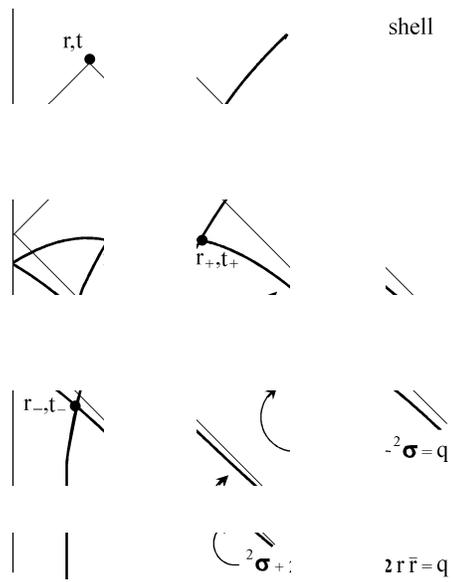
\item[Fig.3.] The world line of the shell crosses the past
hyperboloid of the observation point. For definiteness, the
observation point is shown inside the shell.
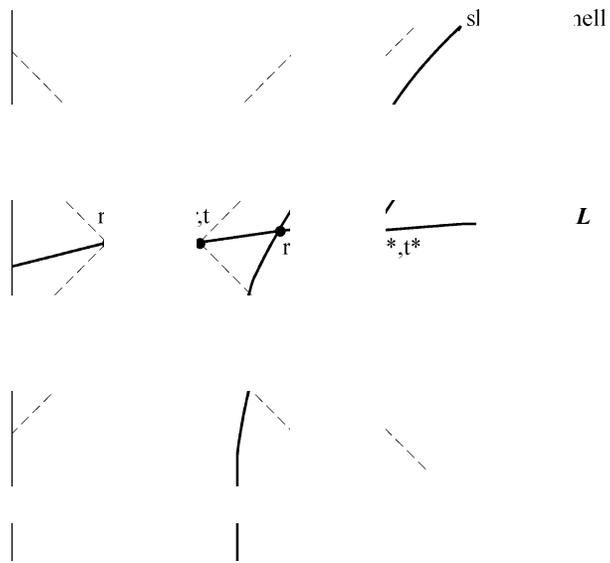
\item[Fig.4.] $L$ is the line specified by Eq. (8.6), and
$r^*,t^*$ is the point at which it crosses the world line of
the shell. The observation point $r,t$ is shown inside the
shell, and the broken lines mark its light cone.
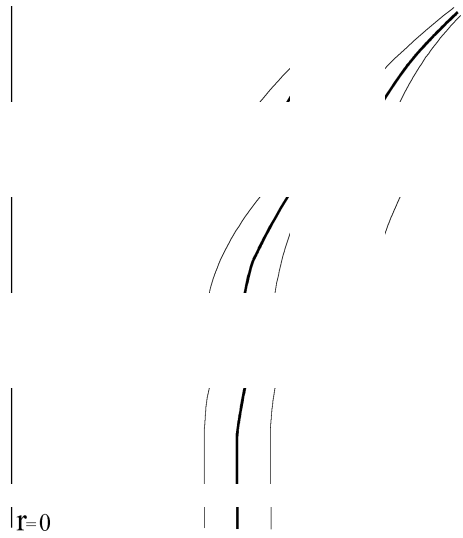
\item[Fig.5.] The band of quantum uncertainty around the world
line of the shell narrows as the speed of expansion increases.
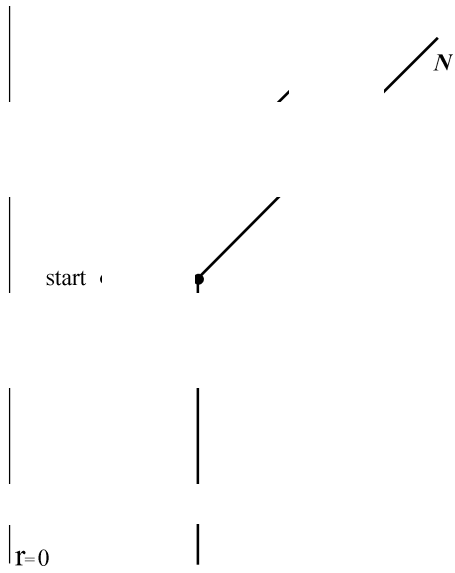
\item[Fig.6.] The world line of the shell at the ultrarelativistic
limit.
\end{itemize}

\newpage
\begin{figure}[p]
\begin{picture}(360,450)
\put(129,215){
\newbox\onebox
\newdimen\onew
\font\onea=onea at 72.27truept
\setbox\onebox=\vbox{\hbox{%
\onea\char0\char1\char2}}
\onew=\wd\onebox
\setbox\onebox=\hbox{\vbox{\hsize=\onew
\parskip=0pt\offinterlineskip\parindent0pt
\hbox{\onea\char0\char1\char2}
\hbox{\onea\char3\char4\char5}
\hbox{\onea\char6\char7\char8}
\hbox{\onea\char9\char10\char11}}}
\ifx\parbox\undefined
    \def\setone{\box\onebox}
\else
    \def\setone{\parbox{\wd\onebox}{\box\onebox}}
\fi
\setone}
\end{picture}
\begin{center}
{\LARGE Fig.1}
\end{center}
\end{figure}
\newpage
\begin{figure}[p]
\begin{picture}(360,450)
\put(87,225){
\newbox\twobox
\newdimen\twow
\font\twoa=twoa at 72.27truept
\setbox\twobox=\vbox{\hbox{%
\twoa\char0\char1\char2\char3}}
\twow=\wd\twobox
\setbox\twobox=\hbox{\vbox{\hsize=\twow
\parskip=0pt\offinterlineskip\parindent0pt
\hbox{\twoa\char0\char1\char2\char3}
\hbox{\twoa\char4\char5\char6\char7}
\hbox{\twoa\char8\char9\char10\char11}
\hbox{\twoa\char12\char13\char14\char15}}}
\ifx\parbox\undefined
    \def\settwo{\box\twobox}
\else
    \def\settwo{\parbox{\wd\twobox}{\box\twobox}}
\fi
\settwo}
\end{picture}
\begin{center}
{\LARGE Fig.2}
\end{center}
\end{figure}
\newpage
\begin{figure}[p]
\begin{picture}(360,450)
\put(107,225){
\newbox\threebox
\newdimen\threew
\font\threea=threea at 72.27truept
\setbox\threebox=\vbox{\hbox{%
\threea\char0\char1\char2}}
\threew=\wd\threebox
\setbox\threebox=\hbox{\vbox{\hsize=\threew
\parskip=0pt\offinterlineskip\parindent0pt
\hbox{\threea\char0\char1\char2}
\hbox{\threea\char3\char4\char5}
\hbox{\threea\char6\char7\char8}
\hbox{\threea\char9\char10\char11}}}
\ifx\parbox\undefined
    \def\setthree{\box\threebox}
\else
    \def\setthree{\parbox{\wd\threebox}{\box\threebox}}
\fi
\setthree}
\end{picture}
\begin{center}
{\LARGE Fig.3}
\end{center}
\end{figure}
\newpage
\begin{figure}[p]
\begin{picture}(360,450)
\put(105,198){
\newbox\fourbox
\newdimen\fourw
\font\foura=foura at 72.27truept
\setbox\fourbox=\vbox{\hbox{%
\foura\char0\char1\char2\char3}}
\fourw=\wd\fourbox
\setbox\fourbox=\hbox{\vbox{\hsize=\fourw
\parskip=0pt\offinterlineskip\parindent0pt
\hbox{\foura\char0\char1\char2\char3}
\hbox{\foura\char4\char5\char6\char7}
\hbox{\foura\char8\char9\char10\char11}
\hbox{\foura\char12\char13\char14\char15}}}
\ifx\parbox\undefined
    \def\setfour{\box\fourbox}
\else
    \def\setfour{\parbox{\wd\fourbox}{\box\fourbox}}
\fi
\setfour}
\end{picture}
\begin{center}
{\LARGE Fig.4}
\end{center}
\end{figure}
\newpage
\begin{figure}[p]
\begin{picture}(360,450)
\put(123,210){
\newbox\fivebox
\newdimen\fivew
\font\fivea=fivea at 72.27truept
\setbox\fivebox=\vbox{\hbox{%
\fivea\char0\char1\char2}}
\fivew=\wd\fivebox
\setbox\fivebox=\hbox{\vbox{\hsize=\fivew
\parskip=0pt\offinterlineskip\parindent0pt
\hbox{\fivea\char0\char1\char2}
\hbox{\fivea\char3\char4\char5}
\hbox{\fivea\char6\char7\char8}
\hbox{\fivea\char9\char10\char11}}}
\ifx\parbox\undefined
    \def\setfive{\box\fivebox}
\else
    \def\setfive{\parbox{\wd\fivebox}{\box\fivebox}}
\fi
\setfive}
\end{picture}
\begin{center}
{\LARGE Fig.5}
\end{center}
\end{figure}
\newpage
\begin{figure}[p]
\begin{picture}(360,450)
\put(129,215){
\newbox\sixbox
\newdimen\sixw
\font\sixa=sixa at 72.27truept
\setbox\sixbox=\vbox{\hbox{%
\sixa\char0\char1\char2}}
\sixw=\wd\sixbox
\setbox\sixbox=\hbox{\vbox{\hsize=\sixw
\parskip=0pt\offinterlineskip\parindent0pt
\hbox{\sixa\char0\char1\char2}
\hbox{\sixa\char3\char4\char5}
\hbox{\sixa\char6\char7\char8}
\hbox{\sixa\char9\char10\char11}}}
\ifx\parbox\undefined
    \def\setsix{\box\sixbox}
\else
    \def\setsix{\parbox{\wd\sixbox}{\box\sixbox}}
\fi
\setsix}
\end{picture}
\begin{center}
{\LARGE Fig.6}
\end{center}
\end{figure}

\end{document}